\documentclass[reqno,12pt]{article}
\usepackage[pdftex]{graphics}

\usepackage{jheppub}

\usepackage{hyperref}
\usepackage{mathtools}
\usepackage{enumerate}
\usepackage{ae}
\usepackage[T1]{fontenc}
\usepackage[ansinew]{inputenc}
\usepackage{mathrsfs}
\usepackage[english]{babel}
\definecolor{darkgreen}{rgb}{0,0.55,0}

\newcommand{\comment}[1]{}

\newcommand{\mincludegraphics}{\includegraphics}

\newcommand{\bl}{\color{blue}}
\newcommand{\rd}{\color{red}}

\newcommand{\beq}{\begin{equation}}
\newcommand{\eeq}{\end{equation}}
\newcommand{\beqq}{\begin{equation*}}
\newcommand{\eeqq}{\end{equation*}}
\newcommand\beqa{\begin{eqnarray}}
\newcommand\eeqa{\end{eqnarray}}
\newcommand\beqaa{\begin{eqnarray*}}
\newcommand\eeqaa{\end{eqnarray*}}
\newcommand\bea{\begin{array}}
\newcommand\eea{\end{array}}

\def\XXint#1#2#3{{\setbox0=\hbox{$#1{#2#3}{\int}$}
\vcenter{\hbox{$#2#3$}}\kern-.5\wd0}}

\newcommand{\nn}{\nonumber}

\newcommand{\neqa}{\nonumber\end{eqnarray}}
\newcommand{\la}[1]{\label{#1}}

\newcommand{\eq}[1]{(\ref{#1})}

\newcommand{\B}{\mathcal{B}}

\newcommand{\Tr}{{\rm Tr}}

\renewcommand{\d}{\partial}

\newcommand{\<}{{\langle}}
\renewcommand{\>}{{\rangle}}

\newcommand{\cA}{{\cal A}}
\newcommand{\cB}{{\cal B}}

\newcommand{\re}{\relax{\rm I\kern-.18em R}}

\renewcommand{\sp}{p\hspace{-.40em}/}

\newcommand{\sq}{q\hspace{-.47em}/}

\def\su2{{SU(2)}}

\def\[{\left[}
\def\]{\right]}

\def\({\left(}
\def\){\right)}
\def\[{\left[}
\def\]{\right]}

\def\<{\langle}
\def\>{\rangle}

\def\sG{\,\slash\!\!\!\! G}

\def\pint{-\hskip-0.41cm \int}
\def\i2{\frac{i}{2}}

\def\O{{\mathcal O}}

\def\spi{\relax{\rm \pi\kern-0.5em /}}
\def\sA{\relax{\rm A\kern-0.5em /}}
\def\sp{\relax{\rm p\kern-0.5em /}}
\def\sd{\relax{\rm \d\kern-0.5em /}}
\def\sk{\relax{\rm k\kern-0.5em /}}
\def\sn{\relax{\rm n\kern-0.5em /}}
\def\sl{\relax{\rm l\kern-0.5em /}}
\def\sP{\relax{\rm P\kern-0.7em /}}
\def\sBethe{\relax{\rm \Bethe\kern-0.5em /}}

\def\bu{{\bf u}}

\newcommand{\Blue}[1]{{\color{blue}#1\color{black}}}
\newcommand{\Red}[1]{{\color{red}#1\color{black}}}

\title{Tailoring Three-Point Functions and Integrability III. Classical Tunneling. }

\author[a,b]{Nikolay Gromov,}
\author[c,d]{Amit Sever}
\author[c]{and Pedro Vieira}

\affiliation[a]{King's College London, Department of Mathematics, \\ London WC2R 2LS, UK}
\affiliation[b]{St.Petersburg INP, \\ Gatchina, 188 300, St.Petersburg, Russia}
\affiliation[c]{Perimeter Institute for Theoretical Physics, \\ Waterloo, Ontario N2L 2Y5, Canada}
\affiliation[d]{School of Natural Sciences,\\Institute for Advanced Study, Princeton, NJ 08540, USA.}

\emailAdd{nikgromov@gmail.com}
\emailAdd{amit.sever@gmail.com}
\emailAdd{pedrogvieira@gmail.com}

\abstract{We compute three-point functions between one large classical operator and two large BPS operators at weak coupling. We consider operators made out of the scalars of $\mathcal{N}=4$ SYM, dual to strings moving in the sphere. The three-point function exponentiates and can be thought of as a \textit{classical tunneling} process in which the classical string-like operator decays into two classical BPS states. From an Integrability/Condensed Matter point of view, we simplified inner products of spin chain Bethe states in a classical limit corresponding to long wavelength excitations above the ferromagnetic vacuum. As a by-product we solved a new long-range Ising model in the thermodynamic limit. }

\keywords{Integrability in gauge/gravity dualities, Three-point functions in $\mathcal{N}=4$ SYM, AdS/CFT correspondence}

\arxivnumber{xxxx.xxxx}

\begin{document}
\maketitle

\section{Introduction}

In recent years there has been outstanding progress in the computation of the planar two point functions of single trace operators in $\mathcal{N}=4$ SYM, see \cite{review} for a recent review. Next, one would like to compute all three-point functions. Together with the two point functions, these are the building blocks for bootstrapping higher point correlators.

\begin{figure}[t]
\centering
\def\svgwidth{8cm}
\input{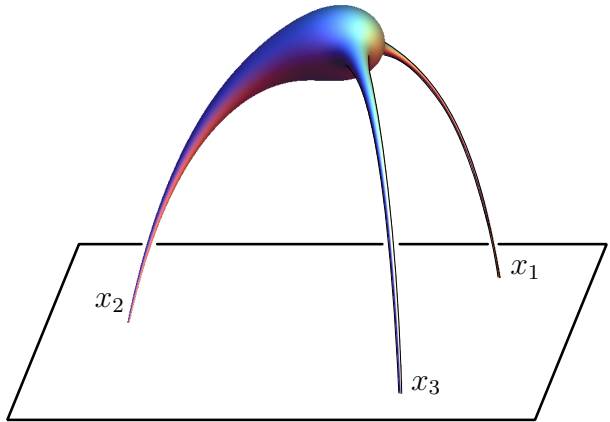}
\caption{A classical extended string decays into two protected classical strings through what we call a \textit{classical tunneling} process. }\label{insect}
\end{figure}

From the string theory side, the problem can be formulated as a calculation of a sphere worldsheet path integral for a string moving in the $AdS_5\times S^5$ background with three (non-normalizable) vertex operator insertions. That path integral is not yet precisely formulated as very little is known about the exact form of the vertex operators. Nevertheless, for two interesting limits there are very promising results at strong coupling. The first limit is when two of the operators are classical (with a large number of fields) and the third operator is a quantum operator (with a small number of fields) \cite{recentpapers1,recentpapers2}, see also \cite{paper2,Roiban:2010fe}.\footnote{This limit can be generalized to higher point functions of two classical operators and several quantum operators \cite{Buchbinder:2010ek,Caetano:2011eb}.} The second limit concerns the recent advances in understanding correlation functions of three classical string states by computing a minimal area surface with three punctures \cite{Janik,Kazama:2011cp}, see also \cite{wavefunctions}.

In this paper we will consider the correlation function of three classical operators \textit{at weak coupling}.
More precisely, we will focus on the case where two of the classical operators are BPS while the third classical operator is dual to an extended classical string. All the operators are made of scalar fields, that is they are dual to strings with non-trivial motion in the sphere. We think of this case as a \textit{classical tunneling process} where the extended non-BPS classical string decays into two BPS classical strings,
(see figure \ref{insect} for an artistic depiction).\footnote{The \textit{classical tunneling} case is conceptually and technically much more involved than the correlation function of two classical strings and one quantum protected string recently studied in \cite{recentpapers2,recentpapers1,Roiban:2010fe,paper2}. A very important difference is that the result for the structure constants $C_{123}$ is exponential for the case of three classical states and is not exponential for the case of two classical operators and one small quantum operator. This is true both at weak and strong coupling. In this paper we will compute the exponent at weak coupling. Of course, it would be very interesting to compute the pre-exponent as well.} \footnote{By {\it classical tunneling} we mean a tunneling process that is controlled by a classical saddle point. }

In the rest of the introduction we present the problem we solve.
\subsection{Formulation of the Problem}
\begin{figure}[t]
\centering
\def\svgwidth{14cm}
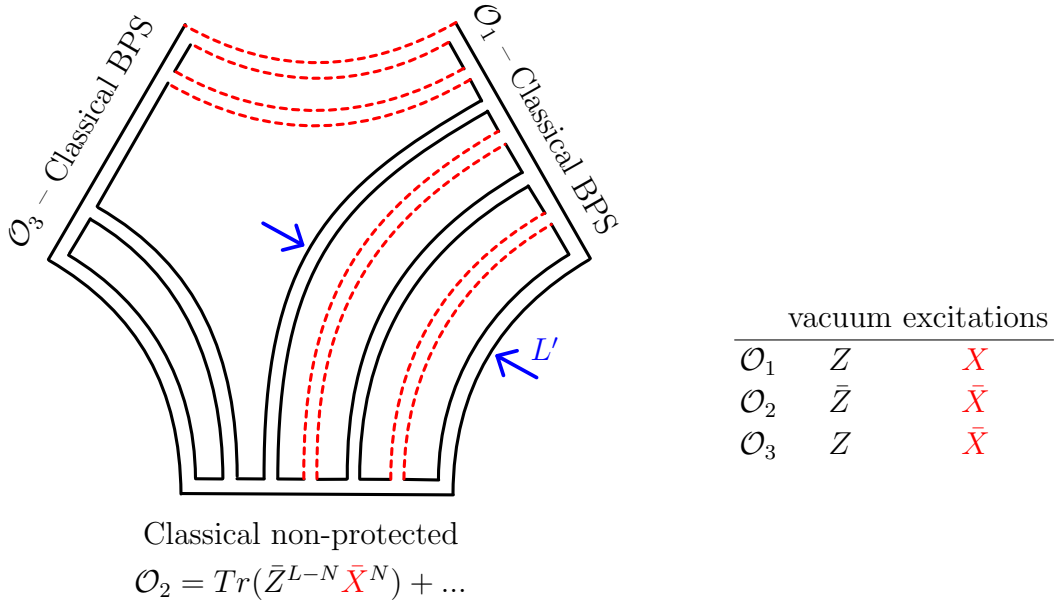
\caption{Three-point function of $SU(2)$ operators at tree level considered in \cite{paper1}.
This is the simplest non-extremal non-trivial configuration. We consider two protected operators $\O_1$ and $\O_3$ and one non-protected operator $\O_2$. All operators are classical, i.e. made of a large number of fields.
 The nontrivial contraction (see arrows in the figure) involves $L'$ fields of $\O_2$. The non-trivial part of the structure constants is given by the universal ratio $\mathcal{A}(\bf u)/\mathcal{B}({\bf u})$, see (\ref{c123disc}). This ratio is in a sense an intrinsic property of the non-protected operator $\O_2$. It only knows about the (length of the) protected operators through $L'$ which only appears in $\mathcal{A}({\bf u})$, see (\ref{AdiscInt}).
}\label{3ptfunction}
\end{figure}
For simplicity we will mostly consider the setup introduced in \cite{paper1} where each of the three operators can be embedded in an $SU(2)$ subsector of the theory, see figure \ref{3ptfunction}. On the other hand, it should be clear that the techniques we will develop can be applied to a much wider set of examples. Some straightforward generalizations are presented in appendix \ref{moreEx} and a more thorough study will be presented elsewhere.

In the setup depicted in figure \ref{3ptfunction}, the operators $\mathcal{O}_1$ and $\mathcal{O}_3$ are classical BPS operators while $\mathcal{O}_2$ is a classical non-BPS operator.
The corresponding structure constant\footnote{The structure constant is well defined once we normalize the two point functions to one. This is implicit throughout the paper.} $C_{123}^{\circ\bullet\circ}$ will depend on the number of each kind of fundamental fields in each of the single trace operators.
It will also depend on the precise form of the operator ${\cal O}^\bullet_2$ which is specified by the Bethe roots ${\bf u}=\{u_i\}$  or equivalently, by the \textit{momenta of the magnons $\bar X$ in the $\bar Z$ ferromagnetic vacuum}, see \cite{paper1} for notation and Bethe ansatz review. In particular, in the limit where all momenta tend to zero $u_j\to \infty$ the structure constant reduces to the structure constant $C_{123}^{\circ\circ\circ}$ of three BPS operators computed in \cite{Lee:1998bxa}.

The tree level result can be written as \cite{paper1}
\footnote{This formula is valid for finite roots $u_j$ quantized according to Bethe ansatz equations.
}
\beq
C_{123}^{\circ\bullet\circ}({\bf u})= C_{123}^{\circ\circ\circ}  \times \[ {\sqrt{{ L \choose N}}}/{{L' \choose N}} \] \times \, \frac{\mathcal{A}({\bf u})}{\mathcal{B}({\bf u})}  \la{c123disc}
\eeq
where $L$ and $N$ are the total length and number of excitation of the non-BPS operator ${\cal O}_2$ while $L'$ is the number of contractions between ${\cal O}_2$ and ${\cal O}_1$. The quantities $\mathcal{A}$ and $\mathcal{B}$ are non-trivial functions of the Bethe roots. They are the main focus of this paper.
The latter is related to the normalization of Bethe eigenstates \cite{Gaudin} and is given by
\beq
\la{BdiscInt} \mathcal{B}({\bf u}) =\sqrt{ \prod_{j\neq i} \frac{u_i-u_j+i}{u_i-u_j} \det_{i,j} \[  \frac{2}{(u_i-u_j)^2+1} +\( \frac{L}{u_i^2+1/4} - \sum_{k} \frac{2}{(u_i-u_k)^2+1}  \)\delta_{ij}  \]}
\eeq
while the former comes from the overlap of part of the non-protected operator $\O_2$ with the protected operator $\O_1$, see figure (\ref{3ptfunction}). In the spin chain language it comes from an overlap of an off-shell Bethe state with a vacuum descendent \cite{paper1}. We have
\beq\la{AdiscInt}
\mathcal{A}({\bf u}) = \sum_{\color{blue} \alpha \color{black} \cup \color{red} \bar\alpha \color{black} = \{u\} } (-1)^{|\color{blue} \alpha \color{black}|}\prod_{\color{blue}  u_a \in \alpha\color{black}}  \(\frac{\color{blue} u_a\color{black}-i/2}{\color{blue} u_a\color{black}+i/2}\)^{L'}  \prod_{\color{blue} u_a\in \alpha\color{black},\color{red} u_b  \in  \bar \alpha\color{black}}  \frac{\color{blue} u_a\color{black}-\color{red} u_b\color{black}+i}{\color{blue} u_a\color{black}-\color{red} u_b\color{black}}
\eeq
The sum in (\ref{AdiscInt}) is over all the ways of partitioning the set of Bethe roots $\{u\}$ into two groups $\alpha$ and $\bar \alpha$. The number of elements in the partition $\alpha$ is denoted as $|\alpha|$.
\begin{figure}[t]
\begin{center}
\mincludegraphics[width=77mm]{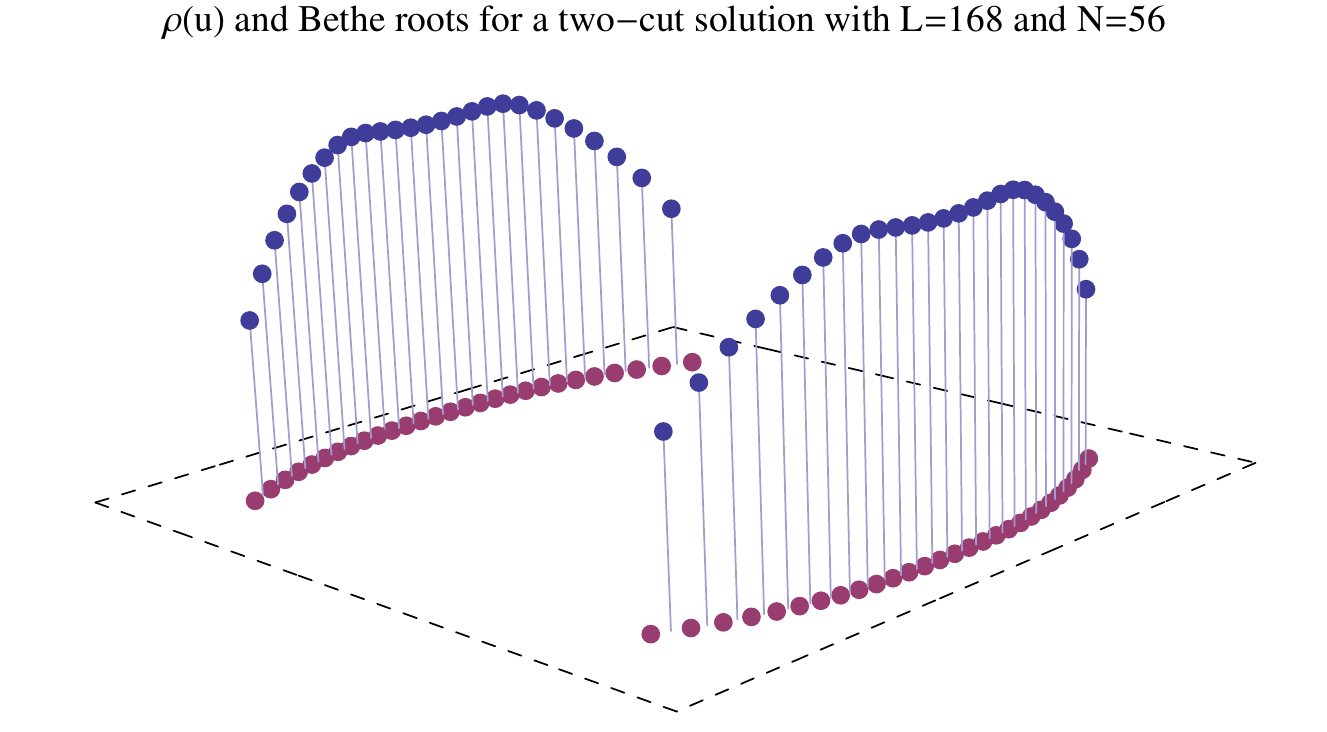}
\mincludegraphics[width=77mm]{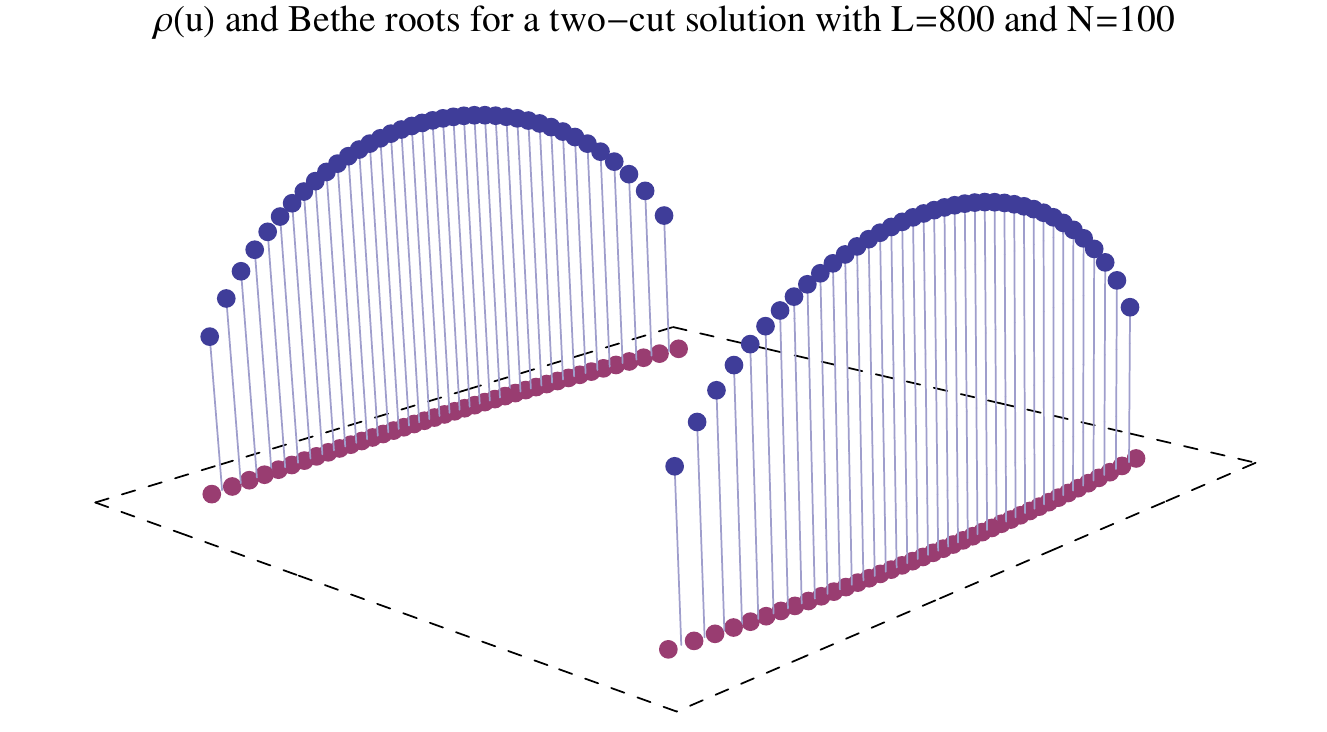}
\end{center}
\caption{Two examples of real configurations of Bethe roots (pink dots in the complex $u$ plane) and corresponding densities $\rho(u)$ (blue dots above the complex $u$ plane). The depicted configurations are the so called two cut solutions. A general classical solution can have an arbitrary large number of cuts. The results presented in the main text are valid for all such configurations.
} \la{figcuts}
\end{figure}

In the classical limit \cite{ing,KMMZ,Sutherland:1995zz},
\beq
u_j \sim N \sim L \to \infty \la{scalinglimit} \,,
\eeq
the Bethe roots condense into smooth cuts described by a density $\rho(u)$, see figure \ref{figcuts}. Hence, in the classical limit, the structure constant will become a functional of this density,
\beq
C_{123}^{\circ\bullet\circ}\(\{u_j\}\)\quad\rightarrow\quad C_{123}^{\circ\bullet\circ}[\rho]\;.
\eeq
It turns out that the continuum limit of (\ref{Bdisc}) and (\ref{Adisc}) is quite a challenging computation, full of remarkable novel structures.
The purpose of the current paper is to perform this continuum limit.

The rest of the paper is organized as follows. In section \ref{Bsec} we study $\mathcal{B}(\bu)$. In section \ref{Asec} we consider $\mathcal{A}(\bu)$. To compute this quantity it turns out that we need to solve a long-range Ising model. This is done in section \ref{secIsing} and appendix \ref{AppPert}. In section \ref{conclusions} we put together $\mathcal{A}(\bu)$ and $\mathcal{B}(\bu)$ to compute $C_{123}^{\circ\bullet\circ}[\rho]$, our main result. In that section we also present several comments of more or less speculative nature. Generalizations to more complicated three-point functions are presented in appendix \ref{moreEx} and some complementary material is contained in appendices \ref{simpleAp}, \ref{detailsSP} and \ref{moretoys}.

\section{Classical Limit of the Norm of Bethe Wave Functions ($\mathcal{B}$)} \la{Bsec}

In this section we initiate our study of the classical scaling limit of Bethe ansatz scalar products arising in (\ref{c123disc}). There are two such products that we need to analyse:  the norm of a Bethe eigenstate (related to ${\cal B}(\bu)$) and  the inner product between an (off-shell) Bethe state and a vacuum descendent (related to $\mathcal{A}(\bu)$). In this section we consider $\mathcal{B}({\bf u})$ given in (\ref{Bdisc}), 
\beq
\la{Bdisc} \mathcal{B}({\bf u}) =\sqrt{ \prod_{j\neq i}^N \frac{u_i-u_j+i}{u_i-u_j} \det M}
\eeq
where
\beq
M_{ij}=   \frac{2}{(u_i-u_j)^2+1} +\( \frac{L}{u_i^2+1/4} - \sum_{k=1}^N \frac{2}{(u_i-u_k)^2+1}  \)\delta_{ij}\;.   \la{Matrix}
\eeq
The goal of this section is to compute this quantity in the classical limit (\ref{scalinglimit}). The final result is given in (\ref{neatB}) below.

\subsection{Derivation} \la{toyBsec}
The determinant in (\ref{Bdisc}) is a determinant of a matrix (\ref{Matrix}) which is a sum of two terms. Let us first drop the second diagonal term, i.e. let us consider the determinant of
\beq
M^{(0)}_{i,j}= \frac{2}{(u_i-u_j)^2+1}\;.
\eeq
\begin{figure}[t]
\begin{center}
\includegraphics[width=100mm]{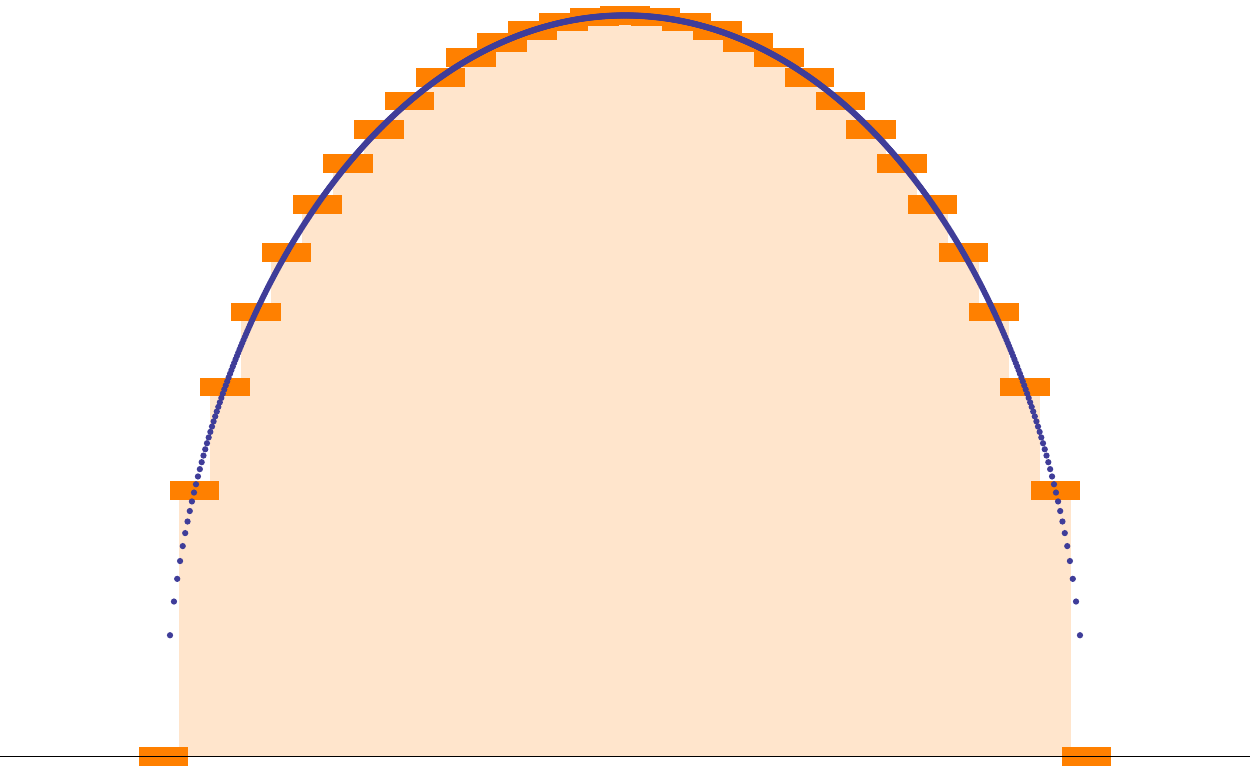}
\end{center}
\caption{In an interval $\(u-du/2,u+du/2\)$ one has $\rho(u) du$ roots described by an approximately constant density $\rho(u)$. As $du \to 0$ the approximation becomes exact. We can use this trick to study exactly quantities which are \textit{roughly diagonal} as described in the main text. A very illustrative example is the toy model (\ref{toyB}).
} \la{blocksfig}
\end{figure}
In the scaling limit (\ref{scalinglimit}), the roots are very large so unless the index $i$ is close to the index $j$ the elements of the matrix can be set to zero. In other words, the matrix is roughly diagonal which
means that the elements of the matrix for which $|i-j|$ is large are suppressed. To see this we split the roots $u_i$ in blocks. These blocks 
contain a large number of roots each but at the same time, within each block, the density $\rho$ can be approximated by a constant, see figure \ref{blocksfig}.
Then inside each block $u_i-u_j=(i-j)/\rho$ and the matrix elements decays as one over a square distance away from the diagonal.
This property of the matrix also allows us to say that the different blocks almost do not interact with each other, that is
\beq
\det M^{(0)}=\det_{ i,j} \frac{2}{(u_i-u_j)^2+1} \simeq \prod_{\text{all blocks}} \det_{ i,j} \frac{2\rho^2}{(i-j)^2+\rho^2} \la{toyB}\;.
\eeq
The range of $i,j$ in the left hand side is $1,\dots,N$ while in the right hand side, for each block, it should in principle
be some large number $\Lambda$, the number of elements in that block, which is at the same time much smaller than $N$.
As we will see, the precise value for $\Lambda\gg 1$ is irrelevant and the result does not depend on the numbers of blocks, provided it is large enough,
which is a consistency requirement on this calculation.
The present analysis leads us to the calculation of the following simpler determinant
\beq
 \det_{  i,j \le \Lambda} \frac{2\rho^2}{(i-j)^2+\rho^2}  \la{toytoyB}
\eeq
for some {\it constant} $\rho$. Since the matrix is translational invariant, it is diagonal in Fourier space. The determinant is simply the product of the corresponding eigenvalues. We have
\beq
  \det_{  i,j \le \Lambda}  \frac{2\rho^2}{(i-j)^2+\rho^2} \simeq \exp\[ \Lambda \int_{0}^1 d\omega \log \lambda(\rho,\omega)  \]  \!\!\!\!\!\!\qquad \text{where}\!\!\! \!\!\!\qquad  \lambda(\rho,\omega) = \sum_{n=-\infty}^{\infty} \frac{2\rho^2}{n^2+\rho^2}  e^{   2 \pi \omega n i}  \,.\la{toyBSol}
\eeq
The sum can be computed exactly,
\beq
\lambda(\rho,\omega) =  \frac{2\pi \rho \cosh\[\pi(1-2\omega) \rho \]}{\sinh(\pi\rho)} \,.
\eeq
This solves the problem (\ref{toytoyB}). The solution to (\ref{toyB}) is now straightforward: we simply multiply over the several blocks
\beq\la{MpreOmega}
\det M^{(0)} \simeq \exp\[ \int_{\mathcal{C}} du\, \rho(u) \int_{0}^1 d\omega \log \lambda(\rho(u),\omega)\]   \,,
\eeq
where $\mathcal{C}=\mathcal{C}_1 \cup \dots \cup \mathcal{C}_K$ are all the cuts along which the roots are distributed and the density of roots on those cuts is $\rho(u)$, see figure \ref{figcuts}.

We now move to the full case, the determinant in (\ref{Bdisc}).
Since we only have to incorporate the term proportional to $\delta_{ij}$ the full matrix is also roughly diagonal, exactly as in our toy model example.
Hence the decomposition into blocks of constant density which we performed above still applies. Let us work out the details.  The new diagonal term in (\ref{Matrix}) contains two pieces: a \textit{source} and a sum. In the classical limit, the source term can be dropped since it is very small
\beq\la{Lpiece}
\frac{L}{u_j^2+1/4} \sim \frac{1}{L}\;.
\eeq
The sum term in the diagonal part of (\ref{Matrix}) is given by
\beq
- \sum_{k} \frac{2}{(u_i-u_k)^2+1}\;.  \la{sum}
\eeq
For $u_k$ far from $u_i$ we can drop the summands in the classical limit (\ref{scalinglimit}). For $u_k$ close to $u_i$ we can use $u_k-u_i=(k-i)/\rho$ where $\rho$ is the density of the distribution of roots around the root $u_i$. That is, $\rho$ is the constant density of roots in the block containing the root $u_i$. Hence
\beq\la{simp}
-\sum_{k} \frac{2}{(u_i-u_k)^2+1} \simeq -\sum_{n} \frac{2 \rho^2}{n^2+\rho^2}  =-2 \pi \rho \coth (\pi \rho)\equiv \alpha(\rho)\;.
\eeq
Altogether, in each block we have
\beq \det_{  i,j \le \Lambda} \( \frac{2\rho^2}{(i-j)^2+\rho^2} +\alpha(\rho) \,\delta_{ij}\)\;.
\eeq
We see that the matrix is simply shifted by a multiple of the  identity matrix.
We conclude that the full determinant in (\ref{Bdisc}) is simply given by a \eq{MpreOmega}
with $\lambda$ shifted to $\alpha$. The integral over $\omega$ can actually be performed analytically.\footnote{We find, up to factors of $i\pi$ (that is signs for $\det M$),
\beqa
\log \det M  &\simeq&   \int_{\mathcal{C}} du\, \rho(u) \int_{0}^1 d\omega
\log  \[ \frac{2\pi \rho(u) \cosh\[\pi(1-2\,\omega))  \rho(u)\color{black} \]}{\sinh[\pi \rho(u)]}
-2 \pi \rho(u) \coth[\pi \rho(u)]\] \nn \\
&=& \int_{\mathcal{C}} \frac{du}{\pi}\, \( \pi\rho(u) \log\[2\pi \rho(u)(e^{2\pi \rho(u)}-1)\]+ {\rm Li}_2\[1-e^{2\pi \rho(u)}\] \)
\eeqa}
The final result can be written in
the following nice form,
\beqa\la{ldM}
\log \det M \simeq \int_{\mathcal{C}} du\,  \[2\int_0^{\rho(u)}\!\!\!d\mu\,\log\[2\sinh(\pi\mu)\]-\rho(u)\log\frac{\sinh[\pi \rho(u)]}{\pi \rho(u)}\]\;.
\eeqa
This concludes the nontrivial part of the computation of ${\cal B}(\bu)$ in (\ref{Bdisc}). There is only one extra simple product multiplying the determinant in (\ref{Bdisc}).
That simple factor is computed in appendix \ref{bextra}. It simply cancels the second term in (\ref{ldM}) so that we end up with the neat final result
\beq
\log{\cal B}\simeq \int\limits_{\mathcal{C}} du\,  \int\limits_0^{\rho(u)} d\mu\,\log\[2\sinh(\pi\mu)\]\;. \la{neatB}
\eeq
Of course, we checked this result against explicit numerics from (\ref{Bdisc}) and found a perfect agreement; see figure \ref{numerics} for similar numerical checks for the more complicated $\mathcal{A}$ which we will discuss in the next section.

Let us end with a small comment. Dropping the source term (\ref{Lpiece}) in (\ref{Matrix}) was a bit dangerous.
Indeed, if we set $L=0$ in (\ref{Matrix}), we remain with a determinant of a matrix where the sum over the element in each row is zero.\footnote{By setting $L$ to zero in (\ref{Matrix}) we mean the coefficient of the source term only, i.e. not changing the position of the Bethe roots.} Hence the matrix has a zero eigenvalue and $\left.\mathcal{B}({\bf u})\right|_{L=0} = 0$. The role of the source term (\ref{Lpiece}) is to regulate that zero eigenvalue. Indeed, a more careful analysis shows the subleading correction to our classical result \eq{neatB} is of order $\log L/L$, coming from
the small eigenvalues.

\section{Inner Product with a Vacuum Descendent ($\mathcal{A}$)} \la{Asec}
We will now move to the more involved computation of the classical limit of (\ref{AdiscInt})
\beq\la{Adisc}
\mathcal{A}({\bf u}) = \sum_{\color{blue} \alpha \color{black} \cup \color{red} \bar\alpha \color{black} = \{u\} } e^{i\tau|\color{blue} \alpha \color{black}|}\prod_{\color{blue}  u_a \in \alpha\color{black}}  \(\frac{\color{blue} u_a\color{black}-i/2}{\color{blue} u_a\color{black}+i/2}\)^{L'}  \prod_{\color{blue} u_a\in \alpha\color{black},\color{red} u_b  \in  \bar \alpha\color{black}}  \frac{\color{blue} u_a\color{black}-\color{red} u_b\color{black}+i}{\color{blue} u_a\color{black}-\color{red} u_b\color{black}}\;.
\eeq
Here we introduced an additional convenient parameter $\tau$ which we call \textit{twist}.\footnote{ It turns out that introducing the twist is quite useful to regulate several expressions. The use of very similar twists to regulate several Bethe ansatz related quantities in the AdS/CFT context can be found e.g. in \cite{twists1,twists2} and in the references therein.} At the end of the day we want to set it to $\pi$.
The main new ingredient in (\ref{Adisc}) is the sum over partitions of the magnons $\{u\}$ into two groups: $\Blue\alpha$ and its complement $\Red{\bar\alpha}$, see figure \ref{partitionsfig}.\footnote{In this section we will often use (redundantly) blue  for quantities related to the partition $\Blue\alpha$ and red for quantities related to $\Red{\bar\alpha}$. All information is contained in a black and white print-out but we believe this section's clarity benefits from being read in a computer or in a color printed copy.} The number of elements in the partition $\Blue\alpha$ is denoted by $\Blue{|\alpha|}$.
We start by considering some simplified toy models to gain some experience and only then move to the actual object of interest.

\begin{figure}[t]
\centering
\def\svgwidth{11cm}
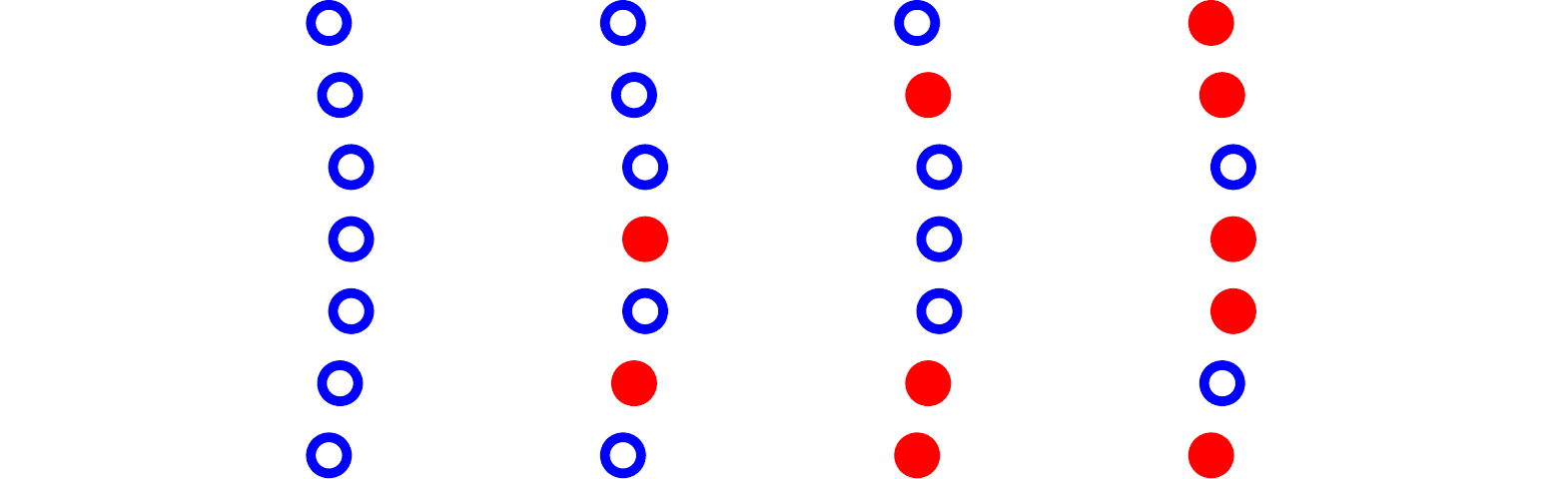
\caption{The second main ingredient in the construction is the function $\mathcal{A}({\bf u})$ in (\ref{Adisc}) graphically depicted in this figure. It is given by a sum over all possible ways of partitioning the Bethe rapidities $\{u_j\}$  into the partitions $\Blue \alpha$ (blue  empty circles) and $\Red{\bar\alpha}$ (filled red circles). For example, for the configuration of seven roots displayed in the figure one sums over the $2^7=128$ way of splitting the Bethe roots into two groups. The purpose of the current section is to address the computation of $\mathcal{A}({\bf u})$ in the classical limit, i.e. when the number of roots becomes very large.
} \la{partitionsfig}
\end{figure}

\subsection{Exactly Solvable Case} \la{toy1sec}
As a first example, lets consider the case where $L'=0$ in $\cA$ (\ref{Adisc}). That is, we consider the classical limit of
\beq
{\cal A}_{L'=0}\equiv \sum_{\color{blue} \alpha \color{black} \cup \color{red} \bar\alpha \color{black} = \{u\} }e^{i\tau|\Blue \alpha|}
\prod_{\scriptsize \begin{array}{c}\Blue{a}\in\Blue{\alpha} \\ \Red{\bar a}\in\Red{\bar\alpha}\end{array}}{\Blue{u_{a}}-\Red{u_{\bar a}}+i\over \Blue{u_{a}}-\Red{u_{\bar a}}} \la{trivial} \,.
\eeq
Curiously enough this case can be solved exactly before even taking the classical limit.
We find that
\beq\la{const}
{\cal A}_{L'=0}=(1+e^{i\tau})^N
\eeq
is independent of the rapidities $\{u\}$! The easiest way to see this is to plug (\ref{trivial}) in \verb"Mathematica" for several values of $N$ and simply check that (\ref{const}) holds indeed.
Also it is not hard to prove this analytically. First note that ${\cal A}_{L'=0}$ is rational with denominator degree that is not less then the numerator degree. The potential single poles of ${\cal A}_{L'=0}$ are where two rapidities, say $u_i$ and $u_j$, coincide. Such pole comes from two type of terms in the sum over partitions: either $\{u_i\in\Blue\alpha,u_j\in\Red{\bar\alpha}\}$ or $\{u_i\in\Red{\bar\alpha},u_j\in\Blue\alpha\}$. For any partition with $\{u_i\in\Blue\alpha,u_j\in\Red{\bar\alpha}\}$, there is a mirror partition that only differs by exchanging $u_i$ and $u_j$. The two contributions to the residue at $u_i=u_j$ only differ by an overall sign (since ${i\over u_i-u_j}=-{i\over u_j-u_i}$). Hence they cancel and the residue is zero. The rational function is therefore a constant! To read the value of that constant (\ref{const}) we can simply send (one by one) all rapidites to infinity.

Even though this example looks trivial it will play a crucial role below.

\subsection{Second Toy Model. Regulated interaction}
As a second toy model example we consider the classical limit of
\beq
{\cal A}_\text{reg}\equiv \sum_{\color{blue} \alpha \color{black} \cup \color{red} \bar\alpha \color{black} = \{u\} }\prod_{\Blue{ a\in\alpha}}e^{i\tau|\Blue \alpha|}\({\Blue{u_{a}}-{i\over2}\over \Blue{u_a}+{i\over2}}\)^{L'}
\prod_{\scriptsize \begin{array}{c}\Blue{a}\in\Blue{\alpha} \\ \Red{\bar a}\in\Red{\bar\alpha}\end{array}}{\Blue{u_{a}}-\Red{u_{\bar a}}+{i\over2}\over \Blue{u_{a}}-\Red{u_{\bar a}}-{i\over2}}\;. \la{T2}
\eeq
The only difference between ${\cal A}_\text{reg}$ and our object of interest ${\cal A}$ is that the interaction ${\Blue{u_{a}}-\Red{u_{\bar a}}+i\over \Blue{u_{a}}-\Red{u_{\bar a}}}$ in  (\ref{Adisc}) is replaced by ${\Blue{u_{a}}-\Red{u_{\bar a}}+{i\over2}\over \Blue{u_{a}}-\Red{u_{\bar a}}-{i\over2}}$ in (\ref{T2}). This might seem like a harmless difference in the classical limit where the rapidities $u_j$ are very large. Indeed, for two well separated roots,
\beq\la{IR}
{\Blue{u_{a}}-\Red{u_{\bar a}}+ i\over \Blue{u_{a}}-\Red{u_{\bar a}} } \simeq  {\Blue{u_{a}}-\Red{u_{\bar a}}+{i\over2}\over \Blue{u_{a}}-\Red{u_{\bar a}}-{i\over2}}\simeq \exp\({i\over \Blue{u_{a}}-\Red{u_{\bar a}}}\)\qquad\text{for}\qquad |\Blue{u_{a}}-\Red{u_{\bar a}}|\gg1
\eeq
Therefore, one may naively think that in the classical limit ${\cal A}_\text{reg}$ and ${\cal A}$ are the same.
However, this is not true. The point is that we also need to consider the interaction in the UV region where the roots are very large but the difference of rapidities is of order $1$ so that (\ref{IR}) can not be used. As we will see below, this UV region is harmless for the toy model ${\cal A}_\text{reg}$ but needs to be taken into account for $\mathcal{A}$. Basically the reason is that the interaction in ${\cal A}_\text{reg}$ is antisymmetric but it is not so for $\mathcal{A}$. See appendix \ref{nonstoc} for some simple illustrative examples of why this makes all the difference.

We will now compute (\ref{T2}) in the classical limit using two different methods: a \textit{direct} one and the \textit{path integral} one.
\subsubsection{Direct Computation}
The direct computation is straightforward. First we notice that
because of the anti-symmetry of the interaction, we can let the product over $\Red{\bar a\in\bar\alpha}$ go over all the roots:
\beq
{\cal A}_\text{reg}\equiv \sum_{\color{blue} \alpha \color{black} \cup \color{red} \bar\alpha \color{black} = \{u\} }\prod_{\Blue{ a\in\alpha}}e^{i\tau|\Blue \alpha|}\({\Blue{u_{a}}-{i\over2}\over \Blue{u_a}+{i\over2}}\)^{L'}
\prod_{\scriptsize \begin{array}{c}\Blue{a}\in\Blue{\alpha} \\ 1\leq b\leq M\end{array}}{\Blue{u_{a}}-{u_{b}}+{i\over2}\over \Blue{u_{a}}-u_{b}-{i\over2}} \;.
\eeq
Next we notice that this sum over partitions can be combined into a single product
\beq\la{overall}
{\cal A}_\text{reg}=\prod_{a=1}^M\[1+e^{i\tau}\({u_a-{i\over2}\over u_a+{i\over2}}\)^{L'}\prod_{b\ne a}^M{u_a-u_b+i/2\over u_a-u_b-i/2}\]\;.
\eeq
This is then straightforward to write in the classical limit by using (\ref{IR}).\footnote{It is very simple to see that there is no contribution from $|u_a -u_b|=O(1)$ because the interaction is anti-symmetric, see
the next section \ref{pathintSec} for more details.} We find
\beq\la{T2classical}
{\cal A}_\text{reg}\simeq\int du\,\rho(u)\,\log\Big[1+\exp (i\sq(u))\Big]
\eeq
with
\beq\la{quasimomenta}
\sq(u)={\tau}-{L'\over u}+\pint dv{\rho(v)\over u-v}
\eeq
where the slash in the integral stands for principal part prescription.

\subsubsection{Path Integral Derivation} \la{pathintSec}
Next, we will re-derive (\ref{T2classical}) by a \textit{path integral} method. We rewrite the sum over partitions in (\ref{T2}) as a path integral over densities $\Blue{\rho_{\alpha}}$ and $\Red{\rho_{\bar\alpha}}$ of particles in the partitions $\Blue\alpha$ and $\Red{\bar \alpha}$ respectively. They are not independent since their sum is fixed to be $\rho$. Hence it is useful to define the single independent variable $\Omega$ as
\beq
\Blue{\rho_{\alpha}}=\frac{\rho+\Omega}{2}\ ,\quad\Red{\rho_{\bar\alpha}}=\frac{\rho-\Omega}{2}\la{defOmega}\;.
\eeq
Then sums over partitions are replaced by
\beq\la{pathint}
\sum_{\Blue\alpha\cup\Red{\bar\alpha} = {\bf u}}  \verb"summand"\,\,\,\,\,\, \longrightarrow  \,\,\,\,\,\,\int [\mathcal{D} \Omega] \, \mu[\Omega]\, \verb"integrand"
\eeq
where
\beq
\log \mu[\Omega]= \int_{\mathcal{C}} du\[\rho\log\rho-\Blue{\rho_{\alpha}}\log{\Blue{\rho_{\alpha}}}-
\Red{\rho_{\bar\alpha}}\log{\Red{\rho_{\bar\alpha}}}\]  \la{entropy}
\eeq
is the usual entropy factor that accounts for the number of microscopic partitions corresponding to a given macroscopic densities $\Blue{\rho_{\alpha}}$ and $\Red{\rho_{\bar\alpha}}$.

In  (\ref{pathint}), the $\verb"integrand"$ appearing in the right hand side is the continuum limit of the $\verb"summand"$ arising in the left hand side. The latter depend on the Bethe roots $\Blue{u_a}$ and $\Red{u_{\bar a}}$ while the former is a functional of the densities  $\Blue{\rho_{\alpha}}$ and $\Red{\rho_{\bar\alpha}}$.
The \verb summand  is
\beq\la{IRsummand}
\verb"summand"=\prod_{\Blue{ a\in\alpha}}e^{i\tau|\Blue \alpha|}\({\Blue{u_{a}}-{i\over2}\over \Blue{u_a}+{i\over2}}\)^{L'}
\prod_{\scriptsize \begin{array}{c}\Blue{a}\in\Blue{\alpha} \\ \Red{\bar a}\in\Red{\bar\alpha}\end{array}}{\Blue{u_{a}}-\Red{u_{\bar a}}+{i\over2}\over \Blue{u_{a}}-\Red{u_{\bar a}}-{i\over2}}
\eeq
which becomes
\beqa\la{IRintegrand}
 \verb"integrand"= \exp \int_{\mathcal{C}} du \,\Blue{\rho_{\alpha}}(u)\[-{i L'\over u}+i \tau+ i  \pint_{\mathcal{C}} dv\,{\Red{\rho_{\bar\alpha}}(v)\over u-v}\]\;.
\eeqa
Putting all the pieces together we find the following path integral representation for our sum,
\beq
{\cal A}_\text{reg} = \int [\mathcal{D} \Omega] \,e^{S[\Omega]} \la{finaltoy}
\eeq
where the \textit{action}
\beq
S[\Omega]= \log \mu[\Omega]+  \log( \verb"integrand")  \la{actiongen}
\eeq
reads
\beq
S[\Omega]= \int_{\mathcal{C}} du\[i\Blue{\rho_{ \alpha}}\sq +\rho\log\rho-\Blue{\rho_{\alpha}}\log{\Blue{\rho_{\alpha}}}-
\Red{\rho_{\bar\alpha}}\log{\Red{\rho_{\bar\alpha}}}\]
\eeq
where the quasi-momenta is given by (\ref{quasimomenta}) and the densities are related to the single independent variable $\Omega(u)$ through (\ref{defOmega}).
We can now compute (\ref{finaltoy})  by saddle point. The saddle point is valid because the action is of order $L$ and $L\to\infty$. The saddle point condition turns out to lead to a very simple  equation for $\Omega$ which can be easily solved leading to
\beq\la{spOmega}
\Omega(u)=\Omega_c(u)=i\rho(u)\tan\frac{\sq(u)}{2}\;.
\eeq
Plugging this into the action leads to
\beq
\log{\cal A}_\text{reg} \simeq S[\Omega_c]=\int_{\mathcal{C}} du\,\rho(u)\log\Big[1+\exp (i\sq(u))\Big]
\eeq
in perfect agreement with our first calculation (\ref{T2classical}).

It is curious to notice the following feature of the path integral derivation: at the saddle point $\Omega(u)$ in (\ref{spOmega}) is purely complex so that the densities $\Blue{\rho_{\alpha}}$ and $\Red{\rho_{\bar\alpha}}$ given by (\ref{defOmega}) are not real.
This is of course a typical feature of a saddle point method: the saddle point can in principle be anywhere in the complex plane.

This concludes our study of the preliminary toy models. A few other interesting simplified examples are presented in appendix \ref{moretoys}. We will now move to the computation of the real object of interest, $\mathcal A(\bu)$.
\subsection{Stochastic Anomaly and the Full $\mathcal{A}$}
To compute $\mathcal A$ using the path integral method introduced above, we have to compute
\beq
\mu[\Omega]\, \verb"integrand"= \left\<e^{i\tau|\color{blue} \alpha \color{black}|}\prod_{\color{blue}  u_a \in \alpha\color{black}}  \(\frac{\color{blue} u_a\color{black}-i/2}{\color{blue} u_a\color{black}+i/2}\)^{L'}  \prod_{\color{blue} u_a\in \alpha\color{black},\color{red} u_b  \in  \bar \alpha\color{black}}  \frac{\color{blue} u_a\color{black}-\color{red} u_b\color{black}+i}{\color{blue} u_a\color{black}-\color{red} u_b\color{black}}\right\>_{\displaystyle\Blue{\rho_{\alpha}},\Red{\rho_{\bar\alpha}}} \la{goal}
\eeq
where the average is over all microscopic configurations $\Blue{\alpha}$ and $\Red{\bar \alpha}$ for fixed macroscopic densities $\Blue{\rho_{\alpha}}$ and $\Red{\rho_{\bar\alpha}}$.
We separate the interaction term into symmetric and antisymmetric parts
by rewriting the $ \verb"integrand"$ as
\beq
\left\<e^{i\tau|\color{blue} \alpha \color{black}|}
\prod_{\color{blue}  u_a \in \alpha\color{black}}  \(\frac{\color{blue} u_a\color{black}-i/2}{\color{blue} u_a\color{black}+i/2}\)^{L'}
\!\!\!\!\!\prod_{\color{blue} u_a\in \alpha\color{black},u_b\neq {\bl u_a} }
 \[
 \frac{\color{blue} u_a\color{black}-u_b\color{black}+i}{\color{blue} u_a\color{black}-u_b\color{black}-i}
 \]^{1/2}
\!\!\!\!\!\prod_{\color{blue} u_a\in \alpha\color{black},\color{red} u_b  \in  \bar \alpha\color{black}}
 \[
 1+\frac{1}{(\color{blue} u_a\color{black}-\color{red} u_b \color{black})^2}
 \]^{1/2}
 \right\>_{\displaystyle\Blue{\rho_{\alpha}},\Red{\rho_{\bar\alpha}}}
 \la{goal2}
\eeq
where the product in the antisymmetric part can be extended to all roots
since the self interaction part gives one in this case.
Note that the first two products can be simplified using the
standard expansion (see previous section and appendix \ref{nonstoc} for more details)
\beq
e^{i\tau|\color{blue} \alpha \color{black}|} \(\frac{\color{blue} u_a\color{black}-i/2}{\color{blue} u_a\color{black}+i/2}\)^{L'}
\prod_{u_b\neq{\bl u_a} }
 \[
 \frac{\color{blue} u_a\color{black}-u_b\color{black}+i}{\color{blue} u_a\color{black}-u_b\color{black}-i}
 \]^{1/2}
\simeq 
\exp\({i\sq({\bl u_a})}\)\la{reso}
\eeq
where $\sq$ is given in (\ref{quasimomenta}).

We notice that all terms in (\ref{goal2}) except for the last one are \textit{not} sensitive to different microscopic realization of the densities.
This means that the weight for two configurations which
differ only by a local exchange of one root in the set $\alpha$
with one root in the set $\bar\alpha$ is the same.\footnote{Indeed from \eq{reso}, the ratio of the weights for the configurations that differ by exchange $u_a$ with $u_{a'}$
gives $\exp\({-{i\sq({u_a})}{}}+{i\sq({u_{a'}})}{}\)\sim 1+c\frac{a-a'}{\rho L}$ where $c\sim 1$ and we use that $u_{a'}=u_{a}+\frac{a'-a}{\rho}$ and that $\d_u\sq\sim 1/L$.}
This allows us to write the first two products in (\ref{goal2}) through the densities
\beq
\mu[\Omega]\, \verb"integrand"=\exp\(i\int\sq {\bl \rho_\alpha} du\)
 \left\<\prod_{\color{blue} u_a\in \alpha\color{black},\color{red} u_b  \in  \bar \alpha\color{black}}
 \[
 1+\frac{1}{(\color{blue} u_a\color{black}-\color{red} u_b \color{black})^2}
 \]^{1/2}
 \right\>_{\displaystyle\Blue{\rho_{\alpha}},\Red{\rho_{\bar\alpha}}}\;.
\eeq
while the last product needs to considered more carefully. 
This remaining average is very local since it goes
to $1$ quite fast. Hence, the nontrivial contribution
comes from the \textit{blocks} of roots such that $u_i-u_j=O(1)$.
We refer the reader to appendix \ref{nonstoc} where we analyze a very similar, but simpler, quantity. What follows mimics closely the treatment of section (\ref{toyBsec}) where we separated the Bethe roots into blocks of constant density, see figure \ref{blocksfig}.
Inside each block we have $u_i-u_j=(i-j)/\rho$. Each block has $\Lambda\gg 1$ roots. Of those, $\frac{\Blue{\rho_{\alpha}}}{\rho} \Lambda$ belong to partition $\Blue\alpha$ and  $\frac{\Red{\rho_{\bar\alpha}}}{\rho} \Lambda$ belong to partition $\Red{\bar\alpha}$. We must sum over which roots are in $\Blue\alpha$ and which roots belong to $\Red{\bar\alpha}$. That is, the contribution of each block is given by $\exp \(\frac{\Lambda}{\rho} \mathcal{Z}\)$ where
\beq
{\cal Z}\equiv \lim_{\Lambda\to \infty} \frac{\rho}{\Lambda} \log \displaystyle\sum_{\{n_i=0,1\}} \prod_{a,b=0}^{\Lambda} \[
1+\frac{\rho^2}{(a-b)^2}\]^{\frac{n_a(1-n_b)}{2}} \la{ising}
\eeq
with
\begin{figure}[t]
\centering
\def\svgwidth{12cm}
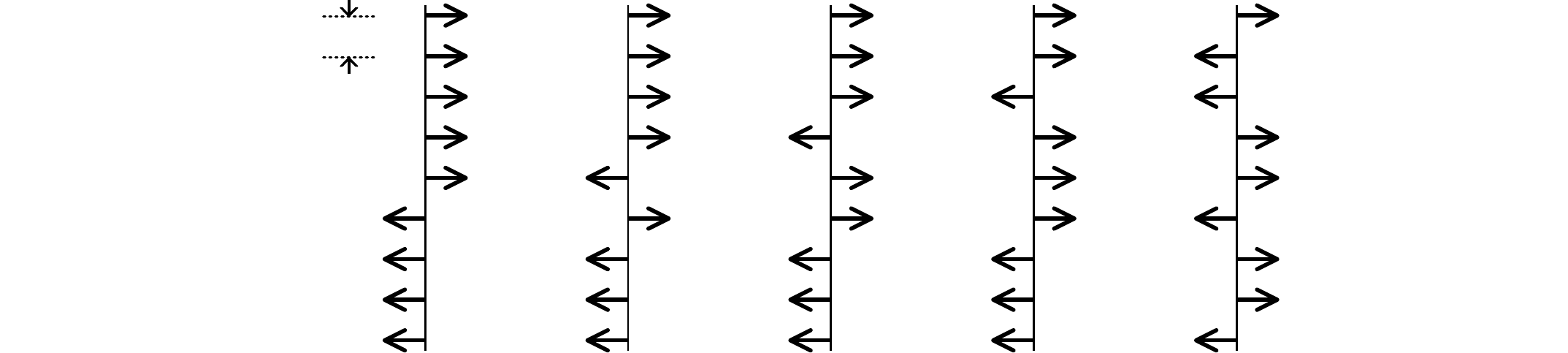
\caption{The effective action is a function of the macroscopic densities $\Blue{\rho_\alpha}$ and $\Red{\rho_{\bar\alpha}}$ only. The interaction between the two set of roots however depends on the microscopic way the roots are distributed into the two groups of fixed macroscopic densities. That dependence result in a contribution to the effective action that we call {\it stochastic anomaly}. It is computed by an Ising model partition function with a long range Coulomb like interaction $\log(1+\rho^2/n^2)$. The figure illustrates the summation of all microscopic spin configurations with a fixed macroscopic density in the Ising model partition function (\ref{ising}).} \la{spinpartitions}
\end{figure}
\beq
\sum_{i=1}^{\Lambda} n_i = \frac{\Blue{\rho_{\alpha}}}{\rho} \Lambda \la{constraint}
\eeq
held fixed. Note that since $\rho=\Blue{\rho_{\alpha}}+ \Red{\rho_{\bar\alpha}}$ we have $\mathcal{Z}=\mathcal{Z}[\Blue{\rho_{\alpha}},\Red{\rho_{\bar\alpha}}]$. We see that to compute this contribution we need to find the partition function of a long-range Ising model (see figure \ref{spinpartitions}). Spin up/down ($n_a=1/n_a=0$) is associated to the partitions $\Blue \alpha$/$\Red{\bar \alpha}$ that contain the root $u_a$, see figure \ref{spinpartitions}. The interaction energy is given by $f(n)\equiv \log(1+\rho^2/n^2)$. Finally note that the Ising model (\ref{ising}) contains automatically the entropy contribution (\ref{entropy}) and reduces to it when $f(n) \to 0$.

\begin{figure}[t]
\begin{center}
\mincludegraphics[width=120mm]{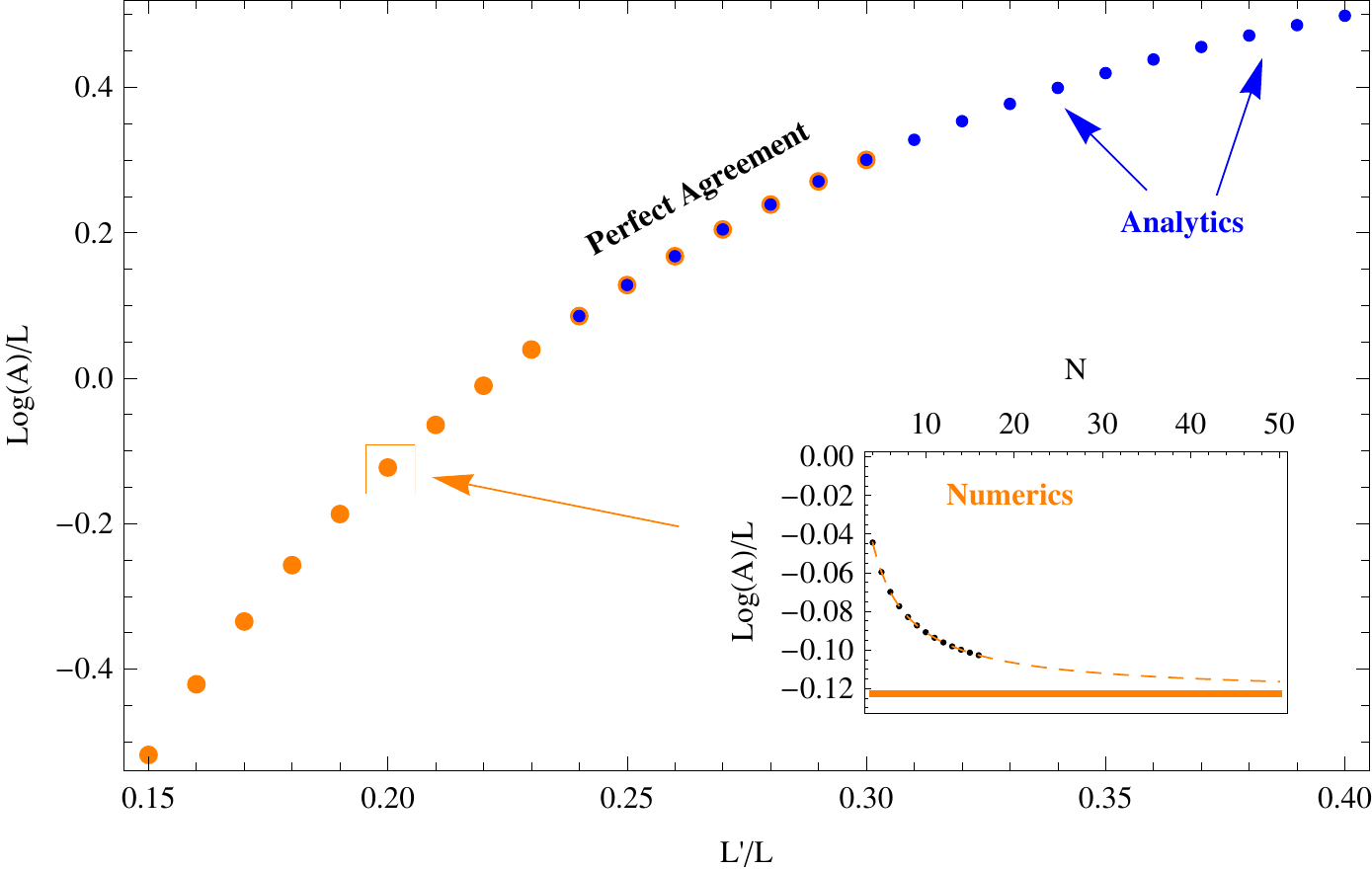}
\end{center}
\caption{To check the main result (\ref{fullA}) we compute $\log\mathcal{A}(\bu)$ from (\ref{Adisc}) for the single cut solution with mode number $n=1$ and filling fraction $\alpha=1/10$  for several $L'/L$. For each $L'$ we compute $\log\mathcal{A}(\bu)$ for more and more Bethe roots (black dots in the bottom right corner). We then fit those points (dashed orange curve in the bottom right corner) to obtain the asymptotic classical value (solid orange curve in the bottom right corner).  The asymptotic results for different $L'/L$ are represented by the orange large dots in the main plot. The small blue dots correspond to the analytic prediction (\ref{fullA}). To make the plot easier to read we represented the analytics and numerics for different (but overlapping) regions. In the overlapping region, the match between the analytics and the numerics is perfect.
} \la{numerics}
\end{figure}

The full path integral action then reads
\beq
S[\Omega]=\int_{\cal C} du \[\mathcal{Z}({\bl \rho_{\alpha}}(u),{\rd \rho_{\bar\alpha}}(u))+i\sq(u){\bl \rho_{\alpha}}(u)\]\la{integrandUV}
\eeq
where (\ref{defOmega}) and (\ref{quasimomenta}).
To conclude the derivation of the full $\mathcal{A}(\bu)$ the main missing ingredient is the computation of the partition function (\ref{ising}). It turns out that computing this partition function is a beautiful but challenging problem. In the next section we will show that
\beq\la{Anom}
\mathcal{Z}({\bl \rho_{\alpha}}(u),{\rd \rho_{\bar\alpha}}(u))={F}(\rho)-{F}(\Blue{\rho_\alpha})-{F}(\Red{\rho_{\bar \alpha}})\;\;,\;\;\quad
{F}(\rho) =\int_0^\rho d\mu \log{\sinh(\pi\mu ) } \,.
\eeq
For now let us take this result and conclude the derivation of  $\mathcal{A}(\bu)$. In fact, at this point, the saddle point computation mimics the end of section \ref{pathintSec} closely. We write the saddle point equation following from (\ref{integrandUV}) and (\ref{Anom}), solve it for $\Omega$, plug it into the action and massage a bit the final result. The details are presented in appendix \ref{detailsSP}. At the end of the day we obtain the very elegant expression,
\beq\la{fullA}
\log {\cal A}\simeq  \oint\limits_{\cup\,{\cal C}_k}{du\over2\pi i}\int\limits_0^{q(u)}d\mu  \,\log(1+e^{i\mu })
\eeq
where the contour integral is taken around the cuts and $q(u)$ is a sort of quasi-momenta, defined in terms of the density by
\beq
q(u)=\tau-{L'\over u}+\int\limits_{\cup\,\mathcal{C}_k} dv{\rho(v)\over u-v}\,. \la{quasiq}
\eeq
The results (\ref{fullA}) and (\ref{neatB}) are the main results of this paper. The fact that (\ref{fullA}) is so beautiful and simple suggests that there should to be a more straightforward derivation.

Since the derivation of the main result (\ref{fullA}) is quite non-trivial it is important to check the final result numerically. A perfect agreement is found, see figure \ref{numerics}. For more details on the single cut solution used for these numerics see \cite{KMMZ} and the appendix C of \cite{paper1}.

\section{Fixing the Anomaly and the Long Range Ising Model} \la{secIsing}
In this section we attack the only missing ingredient in the derivation of the classical limit of $\mathcal{A}(\bu)$, namely the computation of the Ising model partition function (\ref{ising}). The result was anticipated above and is given in (\ref{Anom}).

The derivation is divided into two steps. First, to get inspiration we performed a low density expansion in appendix \ref{AppPert}. The outcome of the perturbative analysis it quite remarkable and iluminating. We find that up to $ \mathcal{O}(\rho^9)$ we have
\beq
\mathcal{Z}({\bl \rho_{\alpha}},{\rd \rho_{\bar\alpha}})={F}(\rho)-{F}(\Blue{\rho_\alpha})-{F}(\Red{\rho_{\bar \alpha}}) \la{decom}
\eeq
with
\beq
F(\rho)= \rho \log \frac{\pi \rho}{e} + \frac{\pi^2 \rho^3}{18}-\frac{\pi^4 \rho^5}{900} + \frac{\pi^6 \rho^7}{19845} + \mathcal{O}(\rho^9)\;. \la{decom2}
\eeq
The structure (\ref{decom}) is highly non-trivial and very inspiring. It leads very naturally to the conjecture that \textit{(\ref{decom}) should hold for finite density}.  Assuming this, we are left with a single function to fix!
This function is then fixed by requiring that the path integral for the exact example (\ref{toy1sec})  gives the correct result as we now explain.

\subsection{Comparing with the Exact Example} \la{comparing}
In this section we will  fix the anomaly (\ref{ising}) which enters in (\ref{integrandUV}). Recall that at this point we are after a single function of a single variable: the function $F(\rho)$ in (\ref{decom}).

The trick is to use the exactly solvable example from  section \ref{toy1sec}. This toy model is related to the full $\mathcal{A}(\bu)$ by setting $L'=0$. The result is given in (\ref{const}),
\beq
\log {\cal A}_{L'=0} = N \log(1+e^{i\tau})\;. \la{simple}
\eeq
In particular it does not depend on the distribution of Bethe roots at all.
On the other hand, if we were to derive it using a path integral we would write down (\ref{integrandUV}) with a very minor simplification, namely $L'=0$. That is
\beq
S=\int\limits_{{\cal C}}du\({i\over2}(\rho+\Omega)(\sG+\tau)+F[\rho]-F[(\rho+\Omega)/2]-F[(\rho-\Omega)/2]\)
\eeq
where
\beq
G(u)=\int\limits_{\cal C} dv{\rho(v)\over u-v}\qquad, \qquad \sG(u)=\pint\limits_{\cal C} dv{\rho(v)\over u-v}\,.
 \la{Gdef}
\eeq
At the saddle point, $S_c=S[\Omega_c]$ should equal (\ref{simple}) for \textit{any} density $\rho$ such that
\beq
\int_{\cal C} \rho(u) = N \,.
\eeq
We see that, from the point of view of the path integral derivation, the trivial result (\ref{simple}) is not obvious at all! It turns out that requiring (\ref{simple}) fixes the anomaly completely as we will now show.

We can simplify our life by choosing a very large imaginary twist $\tau=i t,\;t\to+\infty$. In this limit (\ref{simple}) reduces to
\beq
\log  {\cal A}_{L'=0}  \simeq  N e^{-t}\;. \la{simple2}
\eeq
For large imaginary twists  the factor $e^{i\tau|\Blue \alpha|}$ becomes very small so that the density $\Blue{ \rho_\alpha}$ will be very small. This suppression  leads to major simplifications in the derivation.
For example, it translates into $\Omega=-\rho+2\epsilon$ where $\epsilon$ is very small. Hence
\beq\la{Actep}
S=\int\limits_{\cup\,{\cal C}}du\(-t\epsilon-F(\epsilon)+\epsilon  F'(\rho)+i\epsilon\sG\)
\eeq
and the saddle point equation is simply
\beq
-t+i\sG-F'(\epsilon)+F'(\rho)=0
\eeq
From the result (\ref{decom2}) of the previous section we know that $F(\rho)=\rho\log(\pi\rho/e)+{\cal O}(\rho^3)$.
Using that we deduce from the saddle point equation that
\beq
\epsilon=\frac{1}{\pi}e^{-t}e^{i\sG+F'(\rho)}\;.
\eeq
Finally, on the saddle point, the action \eq{Actep} simplifies to
\beq
S_c=\int\limits_{\cup\,{\cal C}} \epsilon\; du =
\frac{e^{-t}}{\pi}\int\limits_{\cup\,{\cal C}} e^{i\sG+F'(\rho)} du\;.
\eeq
This should be simply equal to (\ref{simple2}). Thus we have to find a function $F$ such
that for {\it any} density $\rho$:
\beq\la{GFeq}
\int\limits_{\cup\,{\cal C}} e^{i\sG+F'(\rho)} du = \pi N\;.
\eeq
This very nontrivial constraint fixes $F(\rho)$ uniquely.

To see this we will perform a small $\rho$ expansion. We use the structure of the perturbative expansion of $F(\rho)$ deduced from our analysis in Appendix \ref{AppPert},
\beq
e^{F'(\rho)} = c_1  \rho + c_2 \rho^3+c_3 \rho^5 + \dots
\eeq
Actually we know the first few terms of this expansion from (\ref{decom2}) but we will not use them. Instead we will derive all the $c_k$ by imposing (\ref{GFeq}).
Expanding the integrand in the left hand side of (\ref{GFeq}) in powers of $\rho$ we get
\beq
\int\limits_{\cup\,{\cal C}} \( c_1\rho+i  c_1 \sG\rho+\[c_2\rho^3-\frac{ c_1\rho\sG^2}{2}\]+\dots\)du=\pi N
\eeq
\begin{itemize}
\item Note that the first term gives perfectly $\pi N$ if we set $c_1=\pi$.
\item All higher orders should give simply zero. To show that
the quadratic term is zero we observe that $4i\pi\sG\rho=G^2(u+i0)-G^2(u-i0)$ and thus
\beq
\int i\pi\sG\rho du=\frac{1}{4}\oint G(u)^2 du = 0
\eeq
where the integration is done by deforming the contour to infinity.
\item
The cubic term can be treated similarly by noticing that $\oint G(u)^3 du =0$ and
$$\frac{G^3(u+i0)-G^3(u-i0)}{12i}=\frac{\pi^3\rho^3}{6}-\frac{\pi\rho\sG^2}{2}\,.$$
This allows us to write
\beq
0=\int\limits_{\cup\,{\cal C}} \(c_2\rho^3-\frac{\pi\rho\sG^2}{2}\)du=
\int\limits_{\cup\,{\cal C}} \(c_2-\frac{\pi^3}{6}\)\rho^3 du
\eeq
since this should hold for an arbitrary $\rho$ we get $c_2=\pi^3/6$.
\end{itemize}
One can continue
to apply this strategy repeatedly fixing all the coefficients in this way. We find
\beq
c_n=\frac{\pi^{2n-1}}{(2n-1)!} \,,
\eeq
which leads to the main result of this section,
\beq
e^{F'(\rho)}=\sinh(\pi\rho)\;. \la{mainresult}
\eeq

It is very easy to check that this is indeed the correct result leading to \eq{GFeq}:
\beq
\int\limits_{\cup\,{\cal C}} e^{i\sG}\sinh(\pi\rho) du
=
\frac{1}{2i}\int\limits_{\cup\,{\cal C}} \(e^{i G(u+i0)}-e^{i G(u-i0)}\) du
=
\frac{1}{2i}\oint\limits_{\cup\,{\cal C}} e^{i G} du
= \pi N\;.
\eeq
At the last step we use that $ G\sim N/u$ for large $u$. The result (\ref{mainresult}) leads to (\ref{Anom}) which was the only missing step in the derivation of the classical limit (\ref{fullA}).

\section{Classical Tunneling and Conclusions} \la{conclusions}
\subsection{Main Result}
We can now put together the results (\ref{neatB}) and (\ref{fullA}) to get the structure constant $C_{123}^{\circ\bullet\circ}$ (\ref{c123disc}) in the classical limit.
This structure constant involves three classical states. Two of them are protected ($\O_1$ and $\O_3$) and one is non-protected ($\O_2$). We call this process \textit{Classical Tunneling}; an artistic depiction is represented in figure \ref{insect}.
In the simplest setup -- represented in figure \ref{3ptfunction} -- all operators are $SU(2)$ operators. For this setup, let us present the final result as a ratio between this structure constant $C_{123}^{\circ\bullet\circ}$ and the known \cite{Lee:1998bxa} structure constant involving three protected operators with the same R-charges $C_{123}^{\circ\circ\circ}$ (times some simple combinatorial factors),\footnote{The non-protected operator $\mathcal{O}_2$ contains $N$ scalars $\bar X$ and $L-N$ scalars $\bar Z$. The number of Wick contractions between operator $\O_2$ and $\O_1$ is $L'$ which is an integer completely fixed by the lengths of the three operators, see figure \ref{3ptfunction}. }
\beq
r \equiv \frac{C_{123}^{\circ\bullet\circ}({\bf u})}{C_{123}^{\circ\circ\circ} {\sqrt{{ L \choose N}}}/{{L' \choose N}} } =\frac{ \mathcal{A}({\bf u})}{\mathcal{B}(\bf u)}\,. \la{rdef}
\eeq
Combining (\ref{neatB}) and (\ref{fullA}) we get a remarkably simple result
\beq
 r =   \exp \int\limits_0^1 dt \[\,\oint\limits_{\cup \,\mathcal{C}_k} \frac{du}{2\pi i}  \,q\log (1-e^{iqt})-   \int\limits_{\cup \,\mathcal{C}_k} du \,\rho\,\log\(2\sinh(\pi t \rho)\)\,\] \la{chapeau} \,.
\eeq
This is the main result of the current paper. Let us summarize the meaning of the several quantities in this formula.  The non-protected operator $\O_2$  is parametrized by a set of $N$ Bethe rapidities $u_j$. In the classical limit they condense into cuts $\mathcal{C}_k$, see figure \ref{figcuts}. The density of roots in these cuts is $\rho(u)$. The number of Wick contractions between operator $\O_2$ and $\O_1$ is $L'$ which is an integer completely fixed by the lengths of the three operators, see figure \ref{3ptfunction}. The quasimomenta $q(u)$ is given in terms of the density and $L'$ as in (\ref{quasiq}).

\subsection{Conclusions and Speculations}
We conclude with a few comments

\begin{itemize}
\item
As discussed in \cite{paper1}, the result (\ref{chapeau}) is highly suggestive. It resembles some sort of phase space integration. Recall that the number of roots in each cut are the action variables of the theory. Our final result is given by a bunch of integrals over the several cuts. Could we shortcut the current derivation by working directly in the action/angle variable basis? Also, the integrand in (\ref{chapeau}) have some kind of fermionic flavor to it. Is our computation dual to some fermionic partition function?
\item
We could also speculate about inspired guesses for the strong coupling value of the structure constants. The idea is that the objects arising in the weak coupling final result (\ref{chapeau}) have a very natural counterpart at strong coupling. For example, the quasi-momenta (\ref{quasiq}) should be related to the eigenvalues of the monodromy matrix constructed from the sigma model Lax pair \cite{KMMZ}. It is therefore very natural to conjecture that (\ref{chapeau}) should describe the strong coupling result up to some very minor modifications of the sort described in \cite{KMMZ,paper1}. Furthermore, in the so called Frolov-Tseytlin limit \cite{FT}, the  strong and weak coupling quasi-momenta are actually the same \cite{KMMZ}. Hence, in this limit, one would expect that no modifications are needed. It would be very interesting to try to check this conjecture.\footnote{At a technical level, it is understood why such match is present for the spectrum problem, see \cite{Kruczenski:2003gt,KMMZ,BKSZ}. At a physical level, there is no good argument why this match should occur and in fact we know that it breaks down at three loops for the spectrum problem. Very recently, it was demonstrated that a match also occurs for three-point functions of two large operators and one light operator \cite{paper2,Georgiou:2011qk}. This match might extend to three-point functions of three classical operators like the one considered in this paper.}  For more speculative discussions on deforming (\ref{chapeau}) to non-zero coupling see \cite{paper1},\cite{paper2}.
\item
Perhaps most interestingly, let us note that the integrals in (\ref{chapeau}) resemble strikingly the strong coupling expressions of \cite{Janik,Kazama:2011cp} obtained using completely different techniques. Unfortunately, these results are still incomplete rendering the comparison very hard.\footnote{For example, \cite{Janik} only takes into account \textit{half} of the full result, the so called \textit{AdS part}; the sphere part is still missing. Also, in \cite{Kazama:2011cp} the contribution from the vertex operator insertions is not accounted.}  It would be extremely interesting to complete these computations and perform a comparison with (\ref{chapeau}).
\item
In this paper we studied the quantities $\mathcal{A}({\bf u})$ and $\mathcal{B}({\bf u})$ in the classical limit.
These quantities are the fundamental building blocks for studying the three-point functions involving one non-protected operator and two protected operators. For simplicity, we focused on the $SU(2)$ setup represented in figure \ref{3ptfunction}. It is simple to generalize this setup to more general configurations as illustrated in appendix \ref{moreEx} and in \cite{ToAppear}. Moreover, recently the structure constant of three non protected SU(2) operators at leading order \cite{paper1} was written in a determinant form \cite{Omar}. It would be interesting to try to use that simplified form for computing the most general structure constants  of non protected operators, see also \cite{ToAppear}.
\item
The classical results we found might be also interesting from a condensed matter perspective. In the condensed matter literature the classical limit (\ref{scalinglimit}) which we consider was first proposed in \cite{Sutherland:1995zz}. It describes long wavelength excitations around the ferromagnetic vacuum. There are infinitely many configurations in this limit, described by any number of cuts as illustrated in figure \ref{figcuts}. We managed to compute two important quantities in this classical limit for all possible such states. The first quantity is the norm of two on-shell Bethe eigenstates,
\beq\la{qt1}
\<0| C(u_1) \dots C(u_N) B(u_1) \dots B(u_N) |0\>\,,
\eeq
where the $u_j$'s obey the Bethe equations in the scaling limit (\ref{scalinglimit}). This quantity is proportional to our  $\mathcal{B}(\bu)$. The second quantity is the inner product of an off-shell Bethe state with a vacuum descendent,
\beq\la{qt2}
\<0| \(S_-\)^N B(u_1) \dots B(u_N) |0\>\,,
\eeq
where the $u_j$'s are unconstrained complex rapidities in the classical limit (\ref{scalinglimit}). This quantity is proportional to our $\mathcal{A}(\bu)$.\footnote{Of course, the precise relation between $\mathcal{B}(\bu)$ and $\mathcal{A}(\bu)$ to (\ref{qt1}) and (\ref{qt2}) depends on the choice of normalization of the Algebraic Bethe ansatz creation and annihilation operators. For precise normalization conventions and precise formulae relating $\mathcal{B}(\bu)$ and $\mathcal{A}(\bu)$ to (\ref{qt1}) and (\ref{qt2}) see \cite{paper1}.}  Inner products like (\ref{qt1}) and (\ref{qt2}) appear abundantly when studying correlation functions and form factors in Integrable models. However, normally, exact results for these quantities are very bulky. We hope that the computations presented in this paper can motivate the study of spin chain correlation functions and form factors in the scaling limit (\ref{scalinglimit}). Could correlation functions in this limit be measurable in ferromagnetic one dimensional chains in the lab?
\item
Finally, let us mention a by-product of our analysis. It turns out that to simplify the computation of the inner product of a Bethe state with a vacuum descendent one needs to solve a very interesting auxiliary problem: a long range Ising model,
\beq
\mathcal{Z}=\lim_{L\to \infty}
\frac{\rho}{L} \log \[\sum_{\{n_i=0,1\}} \exp \( \sum\limits_{a,b=1}^L  n_a \log\[1+\frac{\rho^2}{(a-b)^2}\] (1-n_b)   \)  \]\,.
\eeq
with fixed
\beq
\frac{1}{L}\sum\limits_{a=1}^L  n_a \equiv \frac{\Blue{ \rho_{\alpha}}}{\rho} \,.
\eeq
To our knowledge this model was not considered before.
Quite surprisingly  we found that in the thermodynamic limit where $L\to \infty$ the problem is exactly solvable
\beq
\mathcal{Z} =f(\rho)-f(\Blue{\rho_{ \alpha}})-f(\rho-\Blue{\rho_{ \alpha}}) \qquad \text{with} \qquad f(\rho)=\frac{\text{Li}_2\left(e^{-2 \pi  \rho }\right)}{2 \pi }+\frac{\pi  \rho ^2}{2}-\rho  \log (2)-\frac{\pi }{12}  \nn
\eeq
Technically, this is the most involved computation of the current paper. It would be very interesting to see where this exact result fits in the landscape of integrable models in statistical physics and to explore the physics of this Integrable long range spin chain.
\end{itemize}

\section*{Acknowledgments}

We thank J. Caetano, J. Escobedo, O. Foda, V.Kazakov, R. Janik, J. Toledo, K. Zarembo for  discussions. N.G. would like to thank the Perimeter Institute for warm hospitality. The research of A.S. and P.V. has been supported in part by the Province of Ontario through ERA grant ER 06-02-293. Research at the Perimeter Institute is supported in part by the Government of Canada through NSERC and by the Province of Ontario through MRI. This work was partially funded by the research grants\\ PTDC/FIS/099293/2008 and CERN/FP/109306/2009. This research was supported in part by the National Science Foundation under Grant No. NSF PHY05-51164. This work was supported in part by U.S. Department of Energy grant \#DE-FG02-90ER40542.

\appendix

\section{Simple Products} \la{simpleAp}
\subsection{A Simple Product in $\mathcal{B}$} \la{bextra}
To compute ${\cal B}(\bu)$ (\ref{Bdisc}), we need to add to (\ref{ldM}) the piece
\beq
b_\text{extra}\equiv \log \prod_{j\neq i} \frac{u_i-u_j+i}{u_i-u_j} ={1\over2}\sum_{i\ne j}\log\[1+{1\over(u_i-u_j)^2}\]
\eeq
The computation of this term is pretty much the same as the computation of (\ref{sum}).
We can again split the roots into blocks of constant density, see figure \ref{blocksfig}. Inside each block we can approximate the right hand side by
\beq
  \Lambda \sum_{n=-\Lambda}^{\Lambda} \log\[1+{\rho^2 \over n^2}\] \simeq  \Lambda \log{\sinh[\pi\rho]\over\pi\rho}  \!\!\!\qquad \text{so that}\!\!\! \qquad b_\text{extra}  \,\simeq  \int_{\mathcal{C}} du\,\rho(u)\log{\sinh[\pi\rho(u)]\over\pi\rho(u)} \la{dress}
\eeq
We see that (\ref{dress}) exactly cancels the second term in (\ref{ldM}) so that we end up with the neat final result (\ref{neatB}).
\subsection{Non-Stochastic Anomalies} \la{nonstoc}
Consider the following two products (the product is only over $j$)
\beq
\mathcal{P}_1(u_i) = \prod_{j\neq i}^N \frac{u_i-u_j+i}{u_i-u_j} \qquad, \qquad \mathcal{P}_2(u_i) = \prod_{j\neq i}^N \frac{u_i-u_j+i/2}{u_i-u_j-i/2}
\eeq
where the $u_i \sim N \gg 1$ are distributed according to some smooth density $\rho(u)$. In this classical limit two regions can contribute: A UV region where $u_i,u_j \sim N$ but $|u_i-u_j| =O(1)$ and a IR region where the roots are widely separated. The total result is the product of the two, $$\mathcal{P}_a \simeq \(\mathcal{P}_a\)_\text{IR} \(\mathcal{P}_a\)_\text{UV}\,.$$ The contribution from the IR region is the same for both and simply gives, see (\ref{IR}),
\beq
\(\mathcal{P}_1 \)_\text{IR} = \(\mathcal{P}_2 \)_\text{IR} = \exp \pint \frac{\rho(u)du}{v-u}  \qquad \text{where $v=u_i$}
\eeq
and the slash stands for principal part integration. In the UV region we can use $u_i-u_j= (i-j)/\rho$ to get
\beqa
\(\mathcal{P}_2 \)_\text{UV} &=& \prod_{j=i-\Lambda,j\neq i}^{i+\Lambda} \frac{(i-j)/\rho +i/2}{(i-j)/\rho -i/2}= \prod_{n=-\Lambda,n\neq 0}^{\Lambda} \frac{n +i\rho/2}{n -i\rho/2}=1 \,,
\eeqa
where $\Lambda$ is some large number (but much smaller than $N$) governing the size of the UV region. As we see its precise value is irrelevant. Note that we get no contribution from the UV region simply because the interaction is anti-symmetric. On the other hand,
\beqa
\(\mathcal{P}_1 \)_\text{UV} &=& \prod_{n=-\Lambda,n\neq 0}^{\Lambda} \frac{n +i\rho}{n}=  \frac{\sinh(\pi\rho)}{\pi \rho} \la{short}
\eeqa
We see that the classical limit of $\mathcal{P}_1$ is more subtle from that of $\mathcal{P}_2$. For the latter, we can simply consider the IR region. For the former we need to take into account a short distance contribution (\ref{short}) which we call \textit{non-stochastic anomaly}. Note that to compute this anomaly all we need is the symmetric part of the interaction.\footnote{That is we could very well replace $(u_i-u_j+i)/(u_i-u_j)$ by $\sqrt{1+1/(u_i-u_j)^2}$ when computing the anomaly. We will do precisely this when computing a more complicated anomaly in the main text, called \textit{stochastic anomaly}. }

To compute $\mathcal{A}(u)$ we see that we need to compute a similar but more involved quantity given by (\ref{integrandUV}). There we need to take into account the UV contribution but the roots are not distributed according to smooth distribution as in the previous example. Instead we have to average over all possible distribution of roots into two partitions $\Blue \alpha$ and $\Red{\bar\alpha}$. Hence we denote this contribution as a \textit{stochastic anomaly} to distinguish it from the simplest version above.

There is another closely related quantity called anomaly in the literature \cite{anomalycites},\cite{twists1}  which is roughly\footnote{with $i/2 \to i$ in the definition of $\mathcal{P}_2$} the leading $1/N$ correction to the trivial UV contribution for $\mathcal{P}_2$. The anomalies discussed in this paper are \textit{not} finite size corrections. Instead they are order $1$ effects that can never be discarded. This is a major difference compared to the anomalies in \cite{anomalycites},\cite{twists1}.

\section{Path Integral Saddle Point Details} \la{detailsSP}
In this section we present some of the details concerning the saddle point computation leading to (\ref{fullA}) from (\ref{integrandUV}).
Using the result for the anomaly (\ref{Anom}) in the action (\ref{integrandUV})  we obtain the path integral action
\beq
S=\int\limits_{\cup\,{\cal C}}du\({i\over2}(\rho+\Omega)\sq+F(\rho)-F((\rho+\Omega)/2)-F((\rho-\Omega)/2)\) \la{actionA}
\eeq
where $\sq$ is given in (\ref{quasimomenta}) and $F(\rho) =\int_0^\rho d\mu\,\log\sinh(\pi\mu)$ comes from the computation of the anomaly. The saddle point equation reads
\beq
i\sq=\log{\sinh[\pi(\rho+\Omega_c)/2]\over\sinh[\pi(\rho-\Omega_c)/2]}\quad\Rightarrow\quad\pi\Omega_c=\log{\cosh[\pi(\rho-i\sq/\pi)/2]\over\cosh[\pi(\rho+i\sq/\pi)/2]}
\eeq
Plugging it back into the action, we get
\beqa\la{Sc}
\log \mathcal{A} \simeq S[\Omega_c]&=&\int\limits_{\cup\,{\cal C}} \frac{du}{2\pi}\[{\rm Li}_2(-e^{i\sq-\pi\rho})-{\rm Li}_2(-e^{i\sq+\pi\rho})\]\\
&=&\oint\limits_{\cup\,{\cal C}} \frac{du}{2\pi}{\rm Li}_2(-e^{iq(u)}) \\
&=&\oint\limits_{\cup\,{\cal C}} \frac{du}{2\pi i} \int\limits_0^{q(u)} d\mu \log(1+e^{i\mu})
\eeqa
which is (\ref{fullA}).

\section{Four Extra Identities}\la{moretoys}
In this appendix we discuss a few more simple  cases related to the quantity (\ref{Adisc}),
\beq\la{AdiscA}
\mathcal{A}(L',\tau|{\bf u}) = \sum_{\color{blue} \alpha \color{black} \cup \color{red} \bar\alpha \color{black} = \{u\} } e^{i\tau|\color{blue} \alpha \color{black}|}\prod_{\color{blue}  u_a \in \alpha\color{black}}  \(\frac{\color{blue} u_a\color{black}-i/2}{\color{blue} u_a\color{black}+i/2}\)^{L'}  \prod_{\color{blue} u_a\in \alpha\color{black},\color{red} u_b  \in  \bar \alpha\color{black}}  \frac{\color{blue} u_a\color{black}-\color{red} u_b\color{black}+i}{\color{blue} u_a\color{black}-\color{red} u_b\color{black}}\;.
\eeq
We have
\beqa
&&\!\!\!\!\!\!\!\! \textbf{\underline{off-shell:}}\qquad \mathcal{A}(L',\pi|{\bf u}) = 0 \,\, , \qquad \text{for $0\le L' < N$} \la{id1} \\
&&\!\!\!\!\!\!\!\! \textbf{\underline{on-shell:}}\qquad \mathcal{A}(L',\tau|{\bf u}) = \mathcal{A}(L-L',-\tau|{\bf u})^* \la{id2}\\
&&\!\!\!\!\!\!\!\! \textbf{\underline{on-shell:}}\qquad \mathcal{A}(L',\pi|{\bf u}) = 0 \,\, , \qquad \text{for $0\le L' < N$ or $L-N< L' \le L$ } \la{id3} \\
&&\!\!\!\!\!\!\!\! \textbf{\underline{on-shell:}}\qquad \mathcal{A}(L,\tau|{\bf u}) = (1+e^{i\tau})^N \la{id4}
\eeqa
The first identity, (\ref{id1}) is valid for any set of self-conjugate complex Bethe roots. The last three identities hold provided the roots $u_j$ obey Bethe equations for length $L$,
\beq
\(\frac{u_j+i/2}{u_j-i/2}\)^L=   \prod_{k \neq j }^N \frac{u_j-u_k+i}{u_j-u_k-i} \,.  \nn
\eeq
In what follows we shall derive/discuss the four identities (\ref{id1})-(\ref{id4}).

We start by their derivation. The first one, (\ref{id1}), is the only identity which is \textit{not} trivial to check. The best is to prove it by exhaustion using \verb"Mathematica" or by using some residue arguments as in section \ref{toy1sec}.  Next (\ref{id2}). For $u_j$ obeying the Bethe equations, we have
\beqa
\mathcal{A}(L',\tau|{\bf u}) &=& \sum_{\color{blue} \alpha \color{black} \cup \color{red} \bar\alpha \color{black} = \{u\} } e^{i\tau|\color{blue} \alpha \color{black}|}\prod_{\color{blue}  u_a \in \alpha\color{black}}
\(\frac{\color{blue} u_a\color{black}-i/2}{\color{blue} u_a\color{black}+i/2}\)^{L'-L}
\underbrace{\prod_{\color{blue}  u_a \in \alpha\color{black}}   \(\frac{\color{blue} u_a\color{black}-i/2}{\color{blue} u_a\color{black}+i/2}\)^{L}  }_{\displaystyle=\prod_{\color{blue}  u_a \in \alpha\color{black},\color{red} u_b  \in  \bar \alpha\color{black}}  \frac{\color{blue} u_a\color{black}-\color{red} u_b\color{black}-i}{\color{blue} u_a\color{black}-\color{red} u_b\color{black}+i}
}
\prod_{\color{blue}  u_a \in \alpha\color{black},\color{red} u_b  \in  \bar \alpha\color{black}}  \frac{\color{blue} u_a\color{black}-\color{red} u_b\color{black}+i}{\color{blue} u_a\color{black}-\color{red} u_b\color{black}}\nn
\eeqa
which is clearly the complex conjugate of $\mathcal{A}(L-L',-\tau|{\bf u})$.\footnote{Recall that the set of complex roots is self-conjugate so that $\{u_j\}=\{u_j\}^*$. In particular, in the sums, we can replace $u_j$ by $u_j^*$ at will.}  The identity (\ref{id3}) is a trivial consequence of the previous two identities (\ref{id2}) and (\ref{id1}) once we note that $\tau=\pi$ and $\tau=-\pi$ are the same for (\ref{AdiscA}). Finally, the last identity (\ref{id4}) follows from (\ref{id2}) and (\ref{const}).

Let us now move to the discussion of the physical meaning of these identities.

\begin{itemize}
\item As discussed in the conclusions, the quantity $\mathcal{A}(L',\pi|\bu)$ is proportional to the inner product between a vacuum descendent and a Bethe state. These states we have $N$ spins down and $L'-N$ spins up. Clearly it does not make sense to have $L'<N$. The first identity (\ref{id1}) is a manifestation of this impossibility.  Note that the inner product interpretation only makes sense when the twist $\tau=\pi$ and indeed we only get a zero result in (\ref{id1}) for this particular value of the twist.
\item The second identity (\ref{id2}) resembles strongly the bosonic duality of \cite{twists1}. Would be interesting to explore this analogy further. This might help elucidate the meaning of the twist $\tau$ used throughout this paper. So far we introduced it as a  mathematical tool.
\item The third identity (\ref{id3}) is quite interesting and simple to understand from the three-point function point of view. Let us think of the three-point function as $\<\mathcal{O}_1 | \hat{\mathcal{O}}_3 | \mathcal{O}_2\>$. That is we think of the operator $\mathcal{O}_3$ as a spin chain operator whose average we compute on the states $\mathcal{O}_1$ and $\mathcal{O}_2$ as in \cite{Roiban:2004va,Okuyama:2004bd}. Now, $\O_2$ is a Bethe eigenstate so it is an $SU(2)$ highest weight state of  spin $(L_2-N_2)/2$. The state $\O_1$ is not an highest weight. Instead, it is in the same multiplet as the spin $L_1/2$ highest weight state $\Tr(Z^{L_1})$.
The operator $\mathcal{O}_3$ needs to be large enough to allow for a transition between these states which are in well separated multiplets. In other words, if $L'$ is too large this means that the (effect of the) operator $\O_3$ is too small and hence the transition amplitude will be zero. A precise analysis would lead us to the conclusion that the three-point function should vanish for $L'>L-N$ which is exactly the  content of the new bound  in (\ref{id3}). Note that Bethe states are only highest weight when the Bethe roots obey Bethe equations which is perfectly consistent with the requirement of on-shellness for (\ref{id3}) in contradistinction with (\ref{id1}).
\end{itemize}

\section{Perturbation Theory for the Ising Model} \la{AppPert}
In this appendix we shall derive (\ref{decom}) and (\ref{decom2}) for the partition function (\ref{ising}),
\beq
{\cal Z}\equiv \lim_{\Lambda\to \infty} \frac{\rho}{\Lambda} \log \displaystyle\sum_{\{n_i=0,1\}} \exp\[\sum\limits_{a,b=0}^{\Lambda}  n_a f(a-b) (1-n_b)\]\,,\la{ising2}
\eeq
with
\beq
f(n)=\frac{1}{2}\log\(1+\frac{\rho^2}{n^2}\)
\eeq
and
\beq
\sum_{i=1}^{\Lambda} n_i = \frac{\Blue{\rho_{\alpha}}}{\rho} \Lambda \la{constraint2}
\eeq
held fixed.
\subsection{Combinatoric Measure from the Partition Function}
To warm up we consider first the case where we set $f\to 0$. The corresponding partition function is denoted by $\mathcal{Z}_0$.
The partition function $\mathcal{Z}_0$ just counts the number of permutations when the ratio of the nodes with $n_a=1$
and the nodes with $n_a=0$ is fixed to
\beq
\theta\equiv\frac{\rho_a}{\rho_b} \,.
\eeq
Of course, the result is a simple binomial coefficient. In fact, this is precisely the entropy contribution leading to (\ref{entropy}). We will now reproduce this known result in a slightly different way by introducing a chemical potential which ensures the correct ratio of $1$'s and $0$'s,
\beq\la{anomaly21}
 \mathcal{Z}_0=\lim_{\Lambda\to\infty}
\frac{\rho}{\Lambda}\log{\int \frac{d\mu}{2\pi}\left\langle e^{\sum_{a} i\mu (n_a-\nu)}\right\rangle} \qquad \text{where} \qquad
\nu\equiv \frac{\rho_a}{\rho}=\frac{\theta}{1+\theta} \,.
\eeq
The advantage of the chemical potential is that now each node becomes independent and we easily get
\beq\la{anomaly22}
 \mathcal{Z}_0=\lim_{\Lambda\to\infty}
\frac{\rho}{\Lambda}\log{\int \frac{d\mu}{2\pi}\( e^{i\mu (1-\nu)}+ e^{i\mu (0-\nu)}\)^\Lambda} \,.
\eeq
We just have to compute the integral over $\mu$ which is saturated by the saddle point $\bar\mu=\frac{1}{i}\log\theta$. The above expression then simplifies to
\beq\la{anomaly23}
 \mathcal{Z}_0=\rho\log\rho-\rho_a\log\rho_a-\rho_b\log\rho_b=F_0(\rho)-F_0(\rho_a)-F_0(\rho_b) \,.
\eeq
where $$F_0(\rho)\equiv \rho\log\frac{\pi\rho}{e}=\int_0^\rho\log(\pi\rho)d\rho\,.$$ This coincides precisely with the expected result (\ref{entropy}). This is also the first term of the perturbative expansion anticipated in (\ref{decom2}). In the next section we perform the perturbative expansion to higher orders.

\subsection{Perturbation Theory}
In this section we compute the partition function \eq{ising2} perturbatively in $\rho$.
It is convenient to subtract from $\mathcal{Z}$ the combinatoric measure $\mathcal{Z}_0$ computed in the previous section,
\beq\la{anomaly2}
\widetilde{\mathcal{Z}}\equiv \mathcal{Z}- \mathcal{Z}_0=
\lim_{\Lambda \to\infty}
\frac{\rho}{\Lambda}\log\frac{\int \frac{d\mu}{2\pi}\left\langle e^{\sum_{a,b}  n_a f(a-b)(1-n_b)+\sum_a i\mu (n_a-\nu)}\right\rangle}{\int \frac{d\mu}{2\pi}\left\langle e^{\sum_a i\mu (n_a-\nu)}\right\rangle} \,.
\eeq
The numerator is simply given by \eq{anomaly21}. We will evaluate the numerator in a low density expansion.
A small density expansion coincides with an expansion in $f$, see (\ref{ising2}).
For example, to leading order, the expansion of the
numerator gives
\beq\la{num}
\sum_{a,b}f(a-b) \int \frac{d\mu}{2\pi}\langle  n_a (1-n_b) e^{\sum_a i\mu (n_a-\nu)}\rangle=
\Lambda\int \frac{d\mu}{2\pi} \sum_{n\neq 0}f(n)e^{i\mu (1-2\nu)}\(e^{i\mu (1-\nu)}+e^{-i\mu\nu}\)^{\Lambda-2}
\eeq
This integrals is saturated by the same saddle point $\bar\mu=\frac{1}{i}\log\theta$ as the denominator so that their ratio gives simply
\beq
\widetilde{\mathcal{Z}}=\frac{\rho}{4}\(1-\frac{\Omega^2}{\rho^2}\)\sum_{n\neq 0} f(n)+{\cal O}(f^2)\;.
\eeq
Repeating this calculation up to  $f^3$ order we get\footnote{We define $f(0)\equiv 0$ to avoid writing numerous $\neq 0$.}
\beqa
\widetilde{\mathcal{Z}}/\rho&=&
\frac{\theta}{(\theta+1)^2}\sum_{n} f(n)
+
\frac{\theta^2}{(\theta+1)^4}\sum_{n} f^2(n)\\
\nn&-&
\frac{2\theta^2}{3(\theta+1)^6}
\((1-\theta)^2\sum_{n} f^3(n)+2\theta\sum_{n_1,n_2}f(n_1)f(n_2)f(n_1-n_2)\)+{\cal O}(f^4)\;.
\eeqa
The sums can be evaluated easily at each order in $\rho$ so that we can get an expansion in powers of $\rho$.\footnote{For example,
\beqa
A(\rho,\Omega)&=&\frac{\pi^2\rho^2}{6}\frac{\theta}{(\theta+1)^2}
-\frac{\pi^4\rho^4}{180}\frac{\theta(1+\theta+\theta^2)}{(\theta+1)^4}
+\frac{\pi^6\rho^6}{2835}\frac{\theta(1+\theta+\theta^2)^2}{(\theta+1)^6}+{\cal O}(\rho^8)\;.
\eeqa
}
We notice that a very revealing structure arises once a derivative in  $\theta$ is taken,
\beqa\nn
\d_\theta\widetilde{\mathcal{Z}}&\simeq&-\frac{\pi^2\rho^3}{6}\frac{\theta-1}{(\theta+1)^3}
+\frac{\pi^4\rho^5}{180}\frac{\theta^3-\theta^2+\theta-1}{(\theta+1)^5}
-\frac{\pi^6\rho^7}{2835}\frac{\theta^5-\theta^4+\theta^3-\theta^2+\theta-1}{(\theta+1)^7}\;.
\eeqa
which strongly suggests that
\beq
\d_\theta\widetilde{\mathcal{Z}}= \sum_{n=1}^\infty a_n\rho^{2n+1}\frac{\theta^{2n}-1}{(\theta+1)^{2n+2}} \,.
\eeq
Integrating back and converting back to the $\rho$ variables this translates into the following structure for $\widetilde{\mathcal{Z}}$ (we use that for $\theta=0$ one should get $\widetilde{\mathcal{Z}}=0$ to integrate):
\beq
\widetilde{\mathcal{Z}}=\sum_{n=1}^\infty \frac{a_n}{2n+1} (\rho_a^{2n+1}+\rho_b^{2n+1}-\rho^{2n+1})=\widetilde F(\rho)-\widetilde F(\rho_a)-\widetilde F(\rho_b)\;. \la{structure}
\eeq
where
\beq
\widetilde F(\rho)=\frac{\pi^2\rho^3}{18}-\frac{\pi^4\rho^5}{900}+\frac{\pi^6\rho^7}{19845}+{\cal O}(\rho^9)\,. \la{form}
\eeq
Combining this with the results of the previous section we obtain (\ref{decom2}).

By far the most important outcome of this analysis is the guess for the structure (\ref{structure}) of the partition function. We see that to solve this long range Ising model we simply need to find a single function $\widetilde F(\rho)$. The precise form (\ref{form}) of the low density perturbative expansion of  this function is not so important. In fact we derive the full function $\widetilde F(\rho)$ in section \ref{comparing} by using a completely different approach. We  find that
\beq
\widetilde F(\rho)=\int_{0}^\rho\log\frac{\sinh(\pi\rho)}{\pi\rho}d\rho\,.
\eeq
which matches precisely -- and very nontrivially -- the perturbative result derived above.

We should emphasize that the structure (\ref{structure}) is highly nontrivial. Any small modification of our interaction like
\beq
f(n)=\frac{1}{2}\log(1+\rho^2/n^2) \to \frac{\beta}{2}\log(1+\rho^2/n^2)  \,\,\text{or}\,\,\frac{1}{2}\log(1+\rho^2/n^2+\gamma \rho^4/n^4)
\eeq
would immediately spoil the structure (\ref{structure}). This is not surprising. Integrable models are often isolated points in the moduli and our Ising model (\ref{ising2}) seems to be no exception.

\section{Generalizations} \la{moreEx}
\begin{figure}[t]
\centering
\def\svgwidth{9cm}
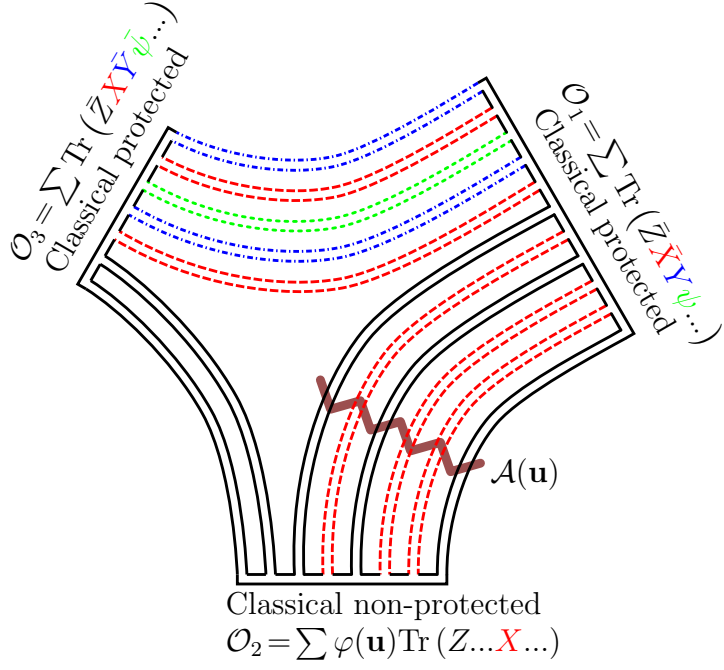
\caption{A tree point function of tree classical operators. Two of the operators ($\O_1$ and $\O_3$) are protected and may contain SU(3) fields and fermions. The third operator, $\O_2$, is a non protected classical operator in the SU(2) sector. The non trivial parts of such structure constant are the same building blocks studied in this paper, $\cA$ and $\B$, coming from the normalization of $\O_2$ and the overlap between $\O_2$ and $\O_1$.}\label{slightly}
\end{figure}

In this paper we studied the quantities $\mathcal{A}_{L'}({\bf u})$ and $\mathcal{B}({\bf u})$ in the classical limit.\footnote{So far we omitted the subscript $L'$ in $\mathcal{A}(\bf u)$ defined as in (\ref{AdiscInt}) since it was always the same. In this section we will always keep this subscript explicit since it will change for the different examples.} The quantities $\mathcal{A}_{L'}({\bf u})$ and $\mathcal{B}({\bf u})$
 are the fundamental building blocks for studying the three-point functions involving one non-protected operator and two protected operators. So far, for simplicity, we considered the $SU(2)$ setup represented in figure \ref{3ptfunction}. It is simple to generalize this setup to more general configurations.

Take for example the setup in figure \ref{slightly}. In this example the protected operators contain all $SU(3)$ scalars and they can even contain fermionic operators. Still, the non-trivial contractions are those between operators $\O_2$ and $\O_1$ and those are the same as in the previous example. Hence, we find that the ratio $r$ defined as in (\ref{rdef}) is given by exactly the same result (\ref{chapeau}),
\beqa
&&r_{\text{fig. \ref{slightly}}} \equiv \frac{C_{123}^{\circ\bullet\circ}({\bf u})}{C_{123}^{\circ\circ\circ}  } =(\text{simple combinatorial factor}) \times \frac{ \mathcal{A}_{L'}({\bf u})}{\mathcal{B}(\bf u)}\\
&& \sim   \exp \int\limits_0^1 dt \[\,\oint\limits_{\,\cup \,\mathcal{C}_k} \frac{du}{2\pi i}  \,q\log (1-e^{iqt})-   \int\limits_{\cup \,\mathcal{C}_k} du \,\rho\,\log\(2\sinh(\pi t \rho)\)\,\] \nn\,.
\eeqa
\begin{figure}[t]
\centering
\def\svgwidth{10cm}
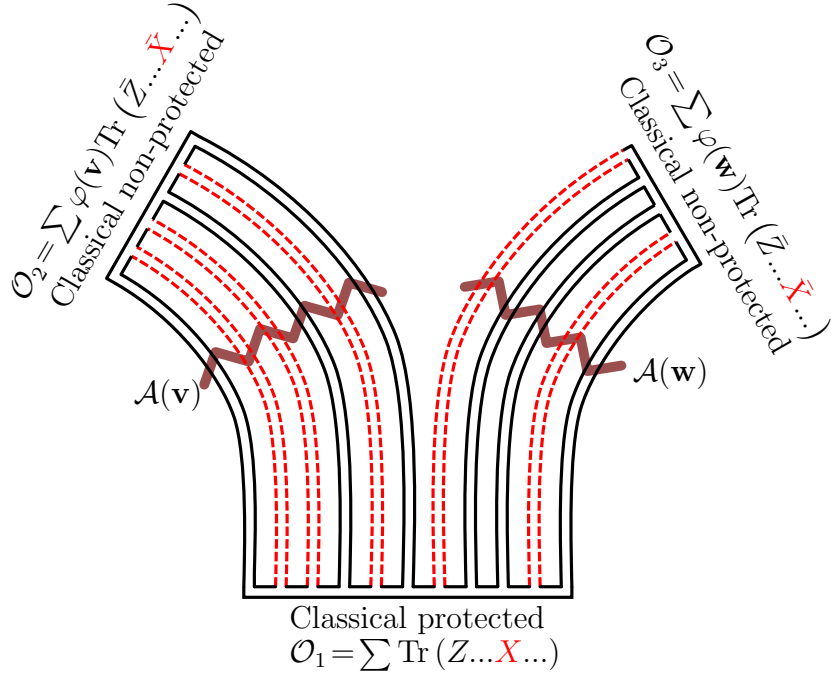
\caption{An extremal correlator of one classical protected operator ($\O_2$) and two non-protected classical operators ($\O_1$ and $\O_3$). The operators $\O_1$ and $\O_3$ are characterized by the Bethe roots $\{v_j\}$ and $\{w_j\}$ respectively. As before, the non trivial parts of such structure constant are the same building blocks studied in this paper, $\cA$ and $\B$, coning from the normalization of $\O_2,\O_3$ and the overlaps between these two operators with parts of $\O_1$.}\label{example2}
\end{figure}
Another example that we can consider in the setup of figure \ref{example2}. In this example we have an extremal correlator between one classical protected operators ($\O_2$) and two classical non protected operators ($\O_1$ and $\O_3$).\footnote{For extremal correlators as the one in figure \ref{example2}, there is also a contribution from the double trace piece of $\O_2$ that contribute at the leading planar order. Here, we only consider the single trace contribution.} There are now two non-trivial group of contractions: the contractions between $\O_2$ and $\O_1$ and those between $\O_2$ and $\O_3$. Each of these contractions corresponds to the inner product between a Bethe state and a vacuum descendent. Let the roots of operator $O_1$ be $\{v_j\}$ and those of operator $\O_3$ be $\{w_j\}$. Those roots will be distributed in cuts as before. We denote the cuts associated to the operator $\O_a$ by $\mathcal{C}_k^{(a)}$. Similarly, the density and quasi-momenta associated to those cuts are, respectively, $\rho^{(a)}$ and $q^{(a)}$.\footnote{The length entering the definition (\ref{quasiq}) of the quasimomenta is $L'$ for $a=3$ and $L-L'$ for $a=1$. }
Then
\beqa
&&r _{\text{fig. \ref{example2}}}\equiv \frac{C_{123}^{\bullet\circ\bullet}({\bf v},{\bf w})}{C_{123}^{\circ\circ\circ}  } =(\text{simple combinatorial factor})  \times \frac{\mathcal{A}_{L-L'}({\bf v})}{\mathcal{B}({\bf v})} \times  \frac{\mathcal{A}_{L'}({\bf w})}{\mathcal{B}({\bf w})} \\
&&\sim  \exp \sum_{a=1,3} \[ \int\limits_0^1 dt\!\! \oint\limits_{\cup \,\mathcal{C}_k^{(a)}} \frac{du}{2\pi i}  \,q^{(a)}\log (1-e^{iq^{(a)}t})-   \int\limits_0^1 dt \!\!\int\limits_{\cup \,\mathcal{C}_k^{(a)}} du \,\rho^{(a)}\,\log\(2\sinh(\pi t \rho^{(a)})\) \]\,. \nn
\eeqa
\begin{figure}[t]
\centering
\def\svgwidth{10cm}
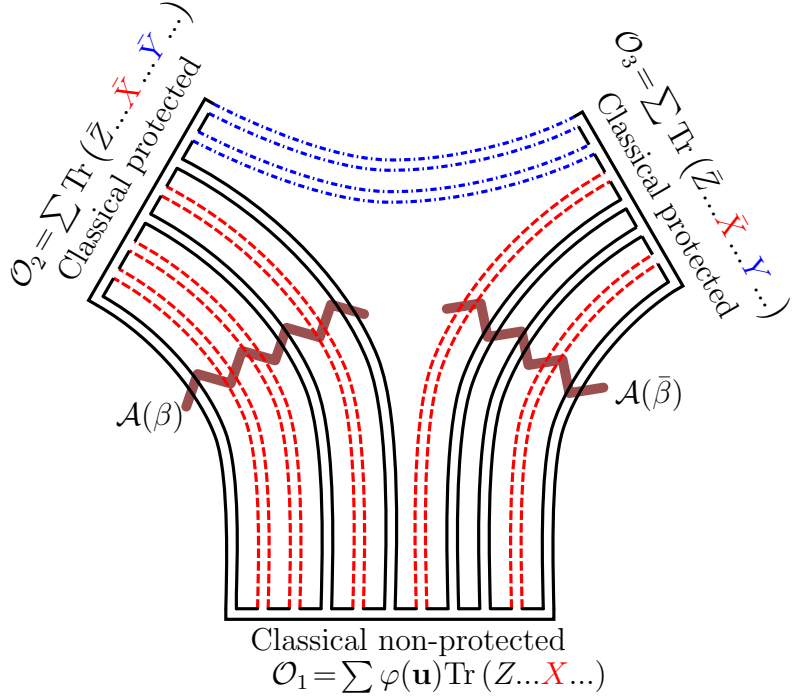
\caption{A tree point function of tree classical SU(2) operators. Two of the operators ($\O_2$ and $\O_3$) are protected and the third operator, $\O_1$, is non protected. For each partition of the roots of $\O_1$ into left and right sets, we have the non trivial overlaps computed by $\cA$ between $\O_2$ and $\O_1$ and between $\O_3$ and $\O_1$.}\label{example3}
\end{figure}
Let us consider one final example, represented in figure \ref{example3}. In this example, there is one non-protected operator $\O_1$ which splits into two protected operators. The two halves of $\O_1$ are non-trivial. This is in contradistinction to all the examples considered so far where all the non-protected operators were split into one trivial half and one non-trivial half, see e.g. figure \ref{3ptfunction}.  The operator $\O_1$ is parametrized by $N$ Bethe roots $\{u_j\}$ which parametrize the $N$ magnon excitations on the corresponding spin chain. When we break that state in two, $N'$ of these magnons will go to the right while $N-N'$ magnons will go to the left. The number of roots $N'$ is fixed by $R$-charge conservation but we still have to sum over which magnons go to the right and which magnons go to the left. That is
\beqa
r_\text{fig. \ref{example3}} &\equiv& \frac{C_{123}^{\bullet\circ\circ}({\bf u})}{C_{123}^{\circ\circ\circ}  } =(\text{simple combinatorial factor}) \sum_{\beta \cup \bar \beta = \bf u\,, \, |\bar\beta|=N'} \frac{\mathcal{A}_{L-L'}(\beta)\mathcal{A}_{L'}(\bar\beta)}{\mathcal{B}(\bf u)} \,. \nn
\eeqa

To summarize, we see that the $\cA$ and $\cB$ are basic blocks that appear in many examples. Their classical limits therefore capture a general pieces of classical correlators. We like to think on that piece as the probability for a macroscopic piece of the classical string to decay into a classical protected string.  We call this process \textit{Classical Tunneling}.


\begin{thebibliography}{99}


\bibitem{review}
  N.~Beisert {\it et al.},
  ``Review of AdS/CFT Integrability: An Overview,''
  arXiv:1012.3982 [hep-th].


\bibitem{recentpapers1}
  K.~Zarembo,
  ``Holographic three-point functions of semiclassical states,''
  JHEP {\bf 1009} (2010) 030
  [arXiv:1008.1059 [hep-th]].


  \bibitem{recentpapers2}
  M.~S.~Costa, R.~Monteiro, J.~E.~Santos and D.~Zoakos,
  ``On three-point correlation functions in the gauge/gravity duality,''
  arXiv:1008.1070 [hep-th].


\bibitem{paper2}
  J.~Escobedo, N.~Gromov, A.~Sever, P.~Vieira,
  ``Tailoring Three-Point Functions and Integrability II. Weak/strong coupling match,''
    [arXiv:1104.5501 [hep-th]].



\bibitem{Roiban:2010fe}
  R.~Roiban and A.~A.~Tseytlin,
  ``On semiclassical computation of 3-point functions of closed string vertex
  operators in $AdS_5 \times S^5$,''
  Phys.\ Rev.\  D {\bf 82}, 106011 (2010)
  [arXiv:1008.4921 [hep-th]].


\bibitem{Buchbinder:2010ek}
  E.~I.~Buchbinder and A.~A.~Tseytlin,
  ``Semiclassical four-point functions in $AdS_5 x S^5$,''
  JHEP {\bf 1102} (2011) 072
  [arXiv:1012.3740 [hep-th]].

\bibitem{Caetano:2011eb}
  J.~Caetano and J.~Escobedo,
  ``On four-point functions and integrability in N=4 SYM: from weak to strong
  coupling,''
  JHEP {\bf 1109}, 080 (2011)
  [arXiv:1107.5580 [hep-th]].



\bibitem{Janik}
R.~A.~Janik, A.~Wereszczynski,
  ``Correlation functions of three heavy operators: The AdS contribution,''
  [arXiv:1109.6262 [hep-th]].

\bibitem{Kazama:2011cp}
  Y.~Kazama, S.~Komatsu,
  ``On holographic three-point functions for GKP strings from integrability,''
  [arXiv:1110.3949 [hep-th]].






\bibitem{wavefunctions}
  R.~A.~Janik, P.~Surowka and A.~Wereszczynski,
  ``On correlation functions of operators dual to classical spinning string states,''
  JHEP {\bf 1005} (2010) 030
  [arXiv:1002.4613 [hep-th]].



\bibitem{paper1}
  J.~Escobedo, N.~Gromov, A.~Sever and P.~Vieira,
  ``Tailoring Three-Point Functions and Integrability,''
  arXiv:1012.2475 [hep-th].


\bibitem{Lee:1998bxa}
  S.~Lee, S.~Minwalla, M.~Rangamani and N.~Seiberg,
  ``Three-point functions of chiral operators in D = 4, N = 4 SYM at  large
  N,''
  Adv.\ Theor.\ Math.\ Phys.\  {\bf 2}, 697 (1998)
  [arXiv:hep-th/9806074].



  \bibitem{Gaudin}
  M. Gaudin, Journal de Physique {\bf 37}, 1087 (1976). $\bullet$
  B. M. McCoy, T. T. Wu, and M. Gaudin, Phys. Rev. D {\bf 23}, 417 (1981). $\bullet$ V.~E.~Korepin,
  ``Calculation of Norms of Bethe Wave Functions,"
  Commun.\ Math.\ Phys.\  {\bf 86} (1982) 391.


\bibitem{ing}
  N.~Beisert, J.~A.~Minahan, M.~Staudacher and K.~Zarembo,
  ``Stringing spins and spinning strings,''
  JHEP {\bf 0309} (2003) 010
  [arXiv:hep-th/0306139].




\bibitem{KMMZ}
  V.~A.~Kazakov, A.~Marshakov, J.~A.~Minahan and K.~Zarembo,
  ``Classical / quantum integrability in AdS/CFT,''
  JHEP {\bf 0405} (2004) 024
  [arXiv:hep-th/0402207].





\bibitem{Sutherland:1995zz}
  B.~Sutherland,
{``Low-Lying Eigenstates of the One-Dimensional Heisenberg Ferromagnet for any
  Magnetization and Momentum,''}
  Phys.\ Rev.\ Lett.\  {\bf 74} (1995) 816.
$\bullet$  A.~Dhar and B.~Sriram Shastry,
  ``Bloch Walls And Macroscopic String States In Bethe's Solution Of The
  Heisenberg Ferromagnetic Linear Chain,''
  Phys.\ Rev.\ Lett.\  {\bf 85} (2000) 2813.




\bibitem{twists1}
  N.~Gromov and P.~Vieira,
  ``Complete 1-loop test of AdS/CFT,''
  JHEP {\bf 0804} (2008) 046
  [arXiv:0709.3487 [hep-th]].

\bibitem{twists2}
  M.~Staudacher,
  ``Review of AdS/CFT Integrability, Chapter III.1: Bethe Ans\'atze and the
  R-Matrix Formalism,''
  arXiv:1012.3990 [hep-th].


\bibitem{FT}
  S.~Frolov and A.~A.~Tseytlin,
  ``Rotating string solutions: AdS / CFT duality in nonsupersymmetric
  sectors,''
  Phys.\ Lett.\  B {\bf 570} (2003) 96
  [arXiv:hep-th/0306143].



\bibitem{Georgiou:2011qk}
  G.~Georgiou,
  ``SL(2) sector: weak/strong coupling agreement of three-point correlators,''
  JHEP {\bf 1109}, 132 (2011).
  [arXiv:1107.1850 [hep-th]].



\bibitem{Kruczenski:2003gt}
  M.~Kruczenski,
  ``Spin chains and string theory,''
  Phys.\ Rev.\ Lett.\  {\bf 93}, 161602 (2004)
  [arXiv:hep-th/0311203].




\bibitem{BKSZ}
  N.~Beisert, V.~A.~Kazakov, K.~Sakai and K.~Zarembo,
  ``The Algebraic curve of classical superstrings on AdS(5) x S**5,''
  Commun.\ Math.\ Phys.\  {\bf 263}, 659 (2006)  [arXiv:hep-th/0502226]. $\bullet$   N.~Beisert, V.~A.~Kazakov, K.~Sakai and K.~Zarembo,
  ``Complete spectrum of long operators in N=4 SYM at one loop,''
  JHEP {\bf 0507}, 030 (2005)
  [arXiv:hep-th/0503200].




\bibitem{ToAppear}
To appear

\bibitem{Omar}
O.~Foda, ``$\mathcal{N} = 4$ SYM structure constants as determinants", To appear


\bibitem{Roiban:2004va}
  R.~Roiban and A.~Volovich,
  ``Yang-Mills correlation functions from integrable spin chains,''
  JHEP {\bf 0409} (2004) 032
  [arXiv:hep-th/0407140].

\bibitem{Okuyama:2004bd}
  K.~Okuyama, L.~-S.~Tseng,
  ``Three-point functions in N = 4 SYM theory at one-loop,''
  JHEP {\bf 0408 } (2004)  055.
  [hep-th/0404190].



\bibitem{anomalycites}
  N.~Beisert and A.~A.~Tseytlin,
  ``On quantum corrections to spinning strings and Bethe equations,''
  Phys.\ Lett.\  B {\bf 629}, 102 (2005)
  [arXiv:hep-th/0509084].
$\bullet$  S.~Schafer-Nameki, M.~Zamaklar and K.~Zarembo,
   ``Quantum corrections to spinning strings in AdS(5) x S**5 and Bethe  ansatz:
 A comparative study,''
  JHEP {\bf 0509}, 051 (2005)
  [arXiv:hep-th/0507189].
$\bullet$  N.~Beisert, A.~A.~Tseytlin and K.~Zarembo,
   ``Matching quantum strings to quantum spins: One-loop vs. finite-size corrections,''
  Nucl.\ Phys.\  B {\bf 715}, 190 (2005)
  [arXiv:hep-th/0502173].
$\bullet$  R.~Hernandez, E.~Lopez, A.~Perianez and G.~Sierra,
   ``Finite size effects in ferromagnetic spin chains and quantum  corrections
  to classical strings,''
  JHEP {\bf 0506}, 011 (2005)
  [arXiv:hep-th/0502188].
$\bullet$  N.~Beisert and L.~Freyhult,
  ``Fluctuations and energy shifts in the Bethe ansatz,''
  Phys.\ Lett.\  B {\bf 622} (2005) 343
  [arXiv:hep-th/0506243].
$\bullet$  N.~Gromov and V.~Kazakov,
  ``Double scaling and finite size corrections in sl(2) spin chain,''
  Nucl.\ Phys.\  B {\bf 736}, 199 (2006)
  [arXiv:hep-th/0510194].

\end{thebibliography}
\end{document}